\documentclass[conference]{IEEEtran}

\ifCLASSINFOpdf
\else
\fi
\usepackage{tcolorbox}
\usepackage{listings}
\usepackage{xcolor}
\usepackage{algorithm}
\usepackage{algpseudocode}
\usepackage{graphicx}

\usepackage{enumitem}

\usepackage{tikz}

\usepackage{mdframed}
\usepackage{xcolor}
\usepackage{pifont}

\newcommand{\xmark}{\ding{55}}
\definecolor{verylightgray}{gray}{0.97}

\usepackage{multirow}
\usepackage[flushleft]{threeparttable}

\usepackage{tikz}
\usepackage{amsmath}
\usepackage{amssymb}
\usepackage{booktabs}
\usepackage{soul}

\usepackage{microtype}

\usepackage{color, colortbl}
\usepackage{hyperref}
\usepackage{float}
\floatstyle{plaintop}
\restylefloat{table}
\usepackage{xurl}
\usepackage{caption} 
\usepackage{comment}

\newcommand{\shortsectionBf}[1]{\vspace{2.5pt}
\noindent {\bf #1}
}

\usepackage{tabularx}

\usepackage{array}
\usepackage{multirow}
\usepackage{pifont}   

\usepackage{xspace}

\newcommand{\sys}{{\textsc{\small{RISK}}}\xspace}

\begin{document}
%

\title{RISK: Auditing Industrial Control Systems for Too-Late-to-Recover Vulnerabilities}


\author{
Syed Ghazanfar Abbas\\
Purdue University\\
\texttt{abbas4@purdue.edu}
\and
Gang Wang\\
University of Illinois Urbana-Champaign\\
\texttt{gangw@illinois.edu}
\and
Dongyan Xu\\
Purdue University\\
\texttt{dxu@purdue.edu}
}

\maketitle
\begin{abstract}
The security of industrial control systems (ICS) is important. Yet most ICS security efforts focus on the {\em detection} of ICS attacks, with much less attention to the {\em recovery} after detection. In this paper, we address this underexplored area by jointly auditing the detection and recovery of ICS. Specifically, we define the \emph{too-late-to-recover (TLTR) vulnerability}, which allows an attack to drain the available recovery margin before being detected, such that the subsequent recovery procedure will fail to bring the ICS back to a safe state due to the insufficient margin. To audit an ICS for TLTR vulnerabilities, we develop \sys, an automated framework that discovers and validates possible TLTR attack scenarios. \sys holistically models and analyzes—statically and dynamically—the PLC control logic, attack detection policies, recovery procedures, and operational behaviors of an ICS to generate TLTR attack scenarios with concrete attack parameters. We evaluate \sys on three ICS testbeds as well as a real-world fertilizer production plant. Across the three testbeds, a total of 392 TLTR attacks are generated and confirmed, whereas only a small fraction of them can be discovered by existing ICS vetting tools. In the real-world plant, \sys identified a critical TLTR vulnerability which was validated by plant engineers.

\end{abstract}

\section{Introduction}
Industrial Control Systems (ICS) operate safety-critical physical processes such as water treatment, chemical manufacturing, and power generation through a multi-layer control and safety architecture~\cite{cardenas2011attacks,mclaughlin2016cybersecurity, stouffer2011guide}: Programmable logic controllers (PLCs) orchestrate the physical process; detection mechanisms raise alarms when the process deviates from expected behavior, and pre-defined {\em recovery procedures} are executed by operators or automated recovery mechanisms to restore the plant to a safe operating state. If deployed, protection mechanisms, such as Safety Instrumented Systems (SIS) and Emergency Shutdown (ESD) systems, provide an additional layer of process safety.

Prior research on adversarial testing and attack generation has largely focused on discovering attacks that achieve their intended physical impact~\cite{song2024genics,ahmed2025attackllm,das2020anomaly,umer2021attack}. Stealthy attacks pursue the same objective while remaining undetected~\cite{urbina2016limiting}. These works primarily characterize attack success in terms of physical impact and detection. However, in industrial control systems, detection does not conclude the attack defense lifecycle; it triggers deployed recovery procedures in order to restore the plant to a safe operating state.

\begin{figure}[t!]
  \centering
  \includegraphics[width=0.90\columnwidth]{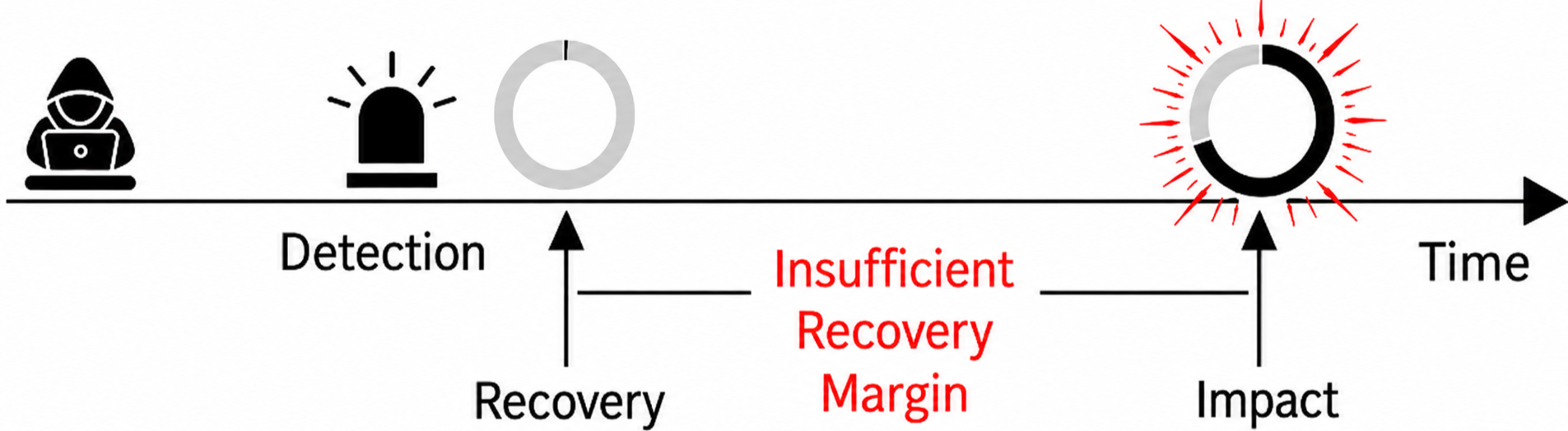}
  \caption{In a TLTR scenario, the available recovery margin is already insufficient upon detection, leading to attack impact before the recovery procedure can successfully finish.}
  \label{fig:RISK Attack}
\end{figure}

Whether such recovery succeeds depends on the state of the physical process when detection occurs. A deployed recovery procedure may be correctly designed and work effectively when sufficient physical and temporal margin remains. However, it may not be able to restore safe system operation, if that margin becomes insufficient upon detection (Figure~\ref{fig:RISK Attack}). Thus, the critical question is not only whether an attack is detected, but whether it is detected while the deployed recovery procedure still has sufficient margin to succeed. 
This motivates a joint ``detection + recovery'' security objective: determining whether the plant remains recoverable at detection time. Unfortunately, existing ICS vetting and attack-generation methods do not explicitly evaluate post-detection recoverability or systematically search for scenarios where recovery margins are insufficient upon detection.

To bridge the gap between detection and recovery, we address the \emph{too-late-to-recover (TLTR) vulnerability} in ICS. We consider an ICS to exhibit a TLTR vulnerability when adversarial manipulations can drain the available physical or temporal recovery margin before detection, such that, once detected, the deployed recovery procedures can no longer restore the plant to its intended safe operating state. We refer to attacks that exploit the TLTR vulnerability \emph{TLTR attacks}. The attacker does not need to evade detection, disable recovery, or exploit an incorrectly designed recovery procedure; it only needs to drive the process to the point, beyond which the deployed recovery procedure will not have sufficient margin to succeed. Depending on the system configuration, exploiting a TLTR vulnerability may result in physical impact, forced shutdown, or other safety-critical consequences.

More specifically, TLTR vulnerabilities can lead to two recovery failure modes. First, recovery can become \emph{infeasible}, where the deployed recovery procedure can no longer restore the process to its intended safe operating state within the applicable recovery constraints. For example, an attacker may reduce safety margins (e.g., the separation between coordinated robots) while remaining within configured limits, such that by the time abnormal behavior is detected, the remaining margin is insufficient to avoid collision (Section~\ref{subsec:recovery-infeasible}). Second, recovery can become \emph{unsafe}, where executing recovery actions triggers cascading violations or hazards, leading to multiple failures (Section~\ref{subsec:recovery-unsafe}).


In this paper, we present \sys, an automated framework for systematically auditing a subject ICS for TLTR vulnerabilities, by discovering and validating possible TLTR attack scenarios.  \sys operates in four steps. First, it extracts process-specific constraints from PLC programs as well as detection policies, recovery procedures, and, where available, protection policies, and organizes them into a unified {\em multi-layer model} that characterizes how the process is governed across the control, detection, and supervisory layers. Second, \sys analyzes operational traces to capture how the process evolves relative to constraints during normal operation, including how quickly the safety margin shrinks and how effective recovery actions are.

Third, \sys uses this model to explore a space of feasible process manipulations that remain within control and safety limits, yet progressively exhaust recovery margins prior to detection. To efficiently navigate this space, \sys leverages a large language model (LLM) agent that suggests candidate TLTR attacks by specifying which process variables to manipulate, when to manipulate them, and by how much. These attacks are evaluated using a virtual PLC (vPLC), a software replica of physical controllers~\cite{gaffurini2024virtual,chowdhury2023case}. Finally, \sys analyzes the resulting execution traces to confirm TLTR attacks, classifying them as recovery-infeasible or recovery-unsafe.

We evaluate \sys on a Fischertechnik manufacturing testbed~\cite{@fp}, a water treatment testbed~\cite{@wp}, and a chemical processing testbed~\cite{@cp}, as well as a real-world fertilizer plant (Case study~\ref{sec:industry}). Most notably, \sys identified a TLTR vulnerability in the plant, which was validated by plant engineers. Across the three testbeds, 392 of 517 validated scenarios generated by \sys (76\%) result in TLTR conditions. Among these cases, 222 are recovery-infeasible scenarios and 170 are recovery-unsafe scenarios. In comparison with existing ICS vetting tools, only 7\% and 8\% of the attacks generated by AttackLLM~\cite{ahmed2025attackllm} and GENICS~\cite{song2024genics}, respectively, are TLTR attacks, indicating that existing solutions are not TLTR-sensitive. We further adopt two state-of-the-art ICS attack detectors, SAIN~\cite{@SAIN} and the physics-based approach by Ghani et al.~\cite{ghaeini}. For SAIN and the physics-based detector, 72\% and 80\% of the detected attacks, respectively, already reach the TLTR condition upon detection and therefore are unrecoverable. 


This paper makes the following contributions:
\begin{itemize}[itemsep = .3mm, topsep = .9mm]
\item \textbf{TLTR vulnerability formulation.} We formulate post-detection recoverability as an {\em integrated} security auditing problem and characterize the TLTR vulnerability as a system condition in which adversarial manipulations drain the recovery margin before detection, leaving the recovery procedures with insufficient margin to restore the process to safe state.



\item \textbf{Joint ``detection + recovery'' ICS auditing.} We present \sys, an automated framework for systematically auditing ICS for TLTR vulnerabilities by discovering and validating possible TLTR attack scenarios. \sys holistically considers the PLC control logic, detection policies, recovery procedures, protection policies, and operational behavior of an ICS to identify recovery-infeasible and recovery-unsafe TLTR conditions.


\item \textbf{Evaluation on multiple platforms.}
We evaluate \sys on three ICS testbeds and a real-world fertilizer plant case study. Across the testbeds, 76\% of validated scenarios generated by \sys result in TLTR conditions, whereas fewer than 10\% of scenarios generated by existing attack-generation methods result in TLTR conditions. In the real-world case study, plant engineers confirmed the TLTR vulnerability found by \sys.


\end{itemize}

\section{Threat Model}
\label{lab:Threatmodel}
\shortsectionBf{System model.} We consider an ICS in which its PLC orchestrates the physical process, its detection mechanisms identify deviations from normal operations, with recovery carried out through deployed recovery procedures. If available, protection mechanisms such as ESD/SIS will enforce additional safety constraints.

\shortsectionBf{ICS operator's perspective.} \sys is an ICS auditing/vetting framework, used by ICS operators and security engineers to systematically identify and validate Too-Late-To-Recover (TLTR) vulnerabilities. As such, we assume access to subject ICS-specific PLC control programs, detection policies, recovery procedures (in documentation), protection policies, and operational traces, which are routinely maintained for ICS operations. The availability of these ICS-specific artifacts is supported by our communications with real-world manufacturers, including the fertilizer factory in our case study (Section~\ref{sec:industry}).

\shortsectionBf{Attacker capabilities.}
We assume that ICS attackers are remote adversaries with access to network-exposed PLCs or supervisory interfaces (e.g., OPC~UA or web interfaces) that permit reading from and writing to a subset of attacker-accessible PLC variables, including sensing variables (e.g., reported measurements), actuation variables (e.g., valve or pump commands), and configuration variables (e.g., setpoints or operating modes). These capabilities have been demonstrated in prior research on ICS attacks~\cite{@SAIN,ironspider}.



The attacker has no physical access and cannot modify PLC control logic, deployed detection mechanisms, operator interfaces, recovery procedures, or safety protection mechanisms. The attacker's actions remain within existing actuation limits, interlocks, and safety constraints, until the detection alarm is raised. Upon detection, we assume that remote access will be disabled to prevent further manipulations. Thus, a TLTR condition must be established by the attack before detection. 


In this paper, we focus on the {\em operators'} perspective to audit the ICS for TLTR vulnerabilities. In practice, adversaries may not have full access to the target ICS. However, we assume the attacker understands the target physical process and its operational behavior through reconnaissance, leveraging publicly accessible information, knowledge, and data about the ICS. Such reconnaissance ability has been observed in real-world ICS attacks~\cite{whitehead2017ukraine}. 

 \section{Recoverability under Adversarial Manipulation}
 
As a primary defense mechanism in an ICS, process recovery aims to restore safe operation upon detection of anomaly/deviation. Its effectiveness relies on a key assumption: when a deviation is detected, sufficient physical and temporal margin remains for corrective actions to succeed. We demonstrate that this assumption can be violated. An adversary can manipulate the process so that, at the time of detection, recovery actions are no longer feasible (Section~\ref{subsec:recovery-infeasible}) or become unsafe to execute (Section~\ref{subsec:recovery-unsafe}). Consequently, the system may no longer be recoverable even though detection and response are triggered as designed.


\subsection{Recovery-Infeasible Scenario: Robot Collision}
\label{subsec:recovery-infeasible}
We illustrate a recovery-infeasible TLTR scenario using the Fischertechnik manufacturing testbed. This example represents a broader class of industrial processes in which successful recovery depends on bounded actuator dynamics and limited physical slack. In such systems, an attacker can progressively narrow the remaining recovery margin until the available recovery actions can no longer restore safe system operation upon detection.



\begin{figure}[t]
  \centering
  \includegraphics[width=.75\linewidth]{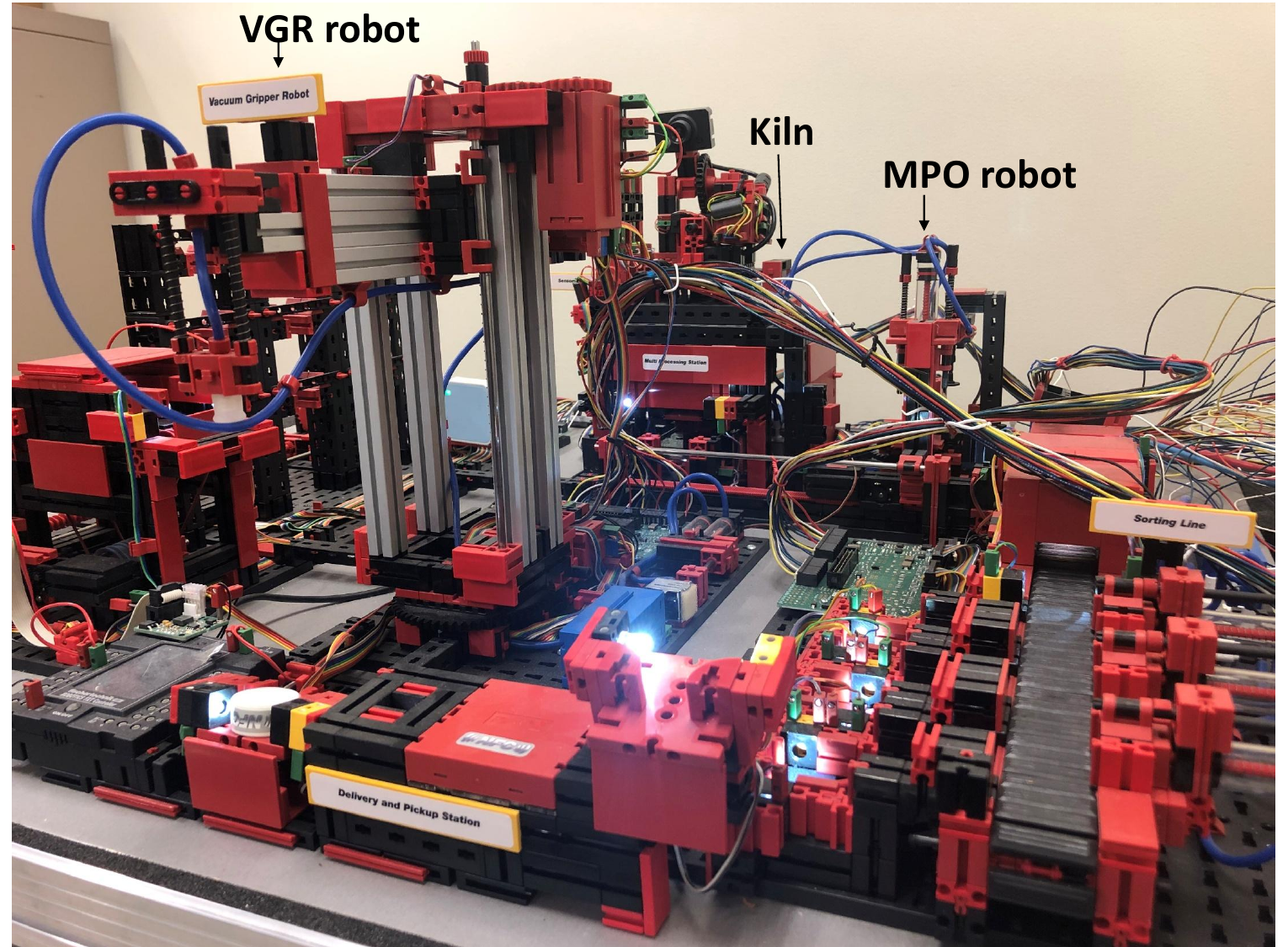}
  \caption{Fischertechnik manufacturing testbed with VGR and MPO robots coordinated to access a shared kiln.
  }
  \label{fig:FTPlant}
\end{figure}

\shortsectionBf{Industrial process.}
The Fischertechnik manufacturing plant~\cite{@fp,sala2023development} consists of two PLC-coordinated robots, a vacuum suction gripper (VGR) and a multi-processing station (MPO), that alternately access a shared kiln (Figure~\ref{fig:FTPlant}). To maximize production throughput, both robots operate with intentionally tight temporal separation while the PLC schedules their access to prevent simultaneous entry into the kiln.

\shortsectionBf{Recovery actions.} Through the HMI, the operator can gradually reduce the robot speed to restore safe operation or initiate an emergency shutdown to stop the robot.

\shortsectionBf{Attack.}
The attacker, with remote access to network-exposed PLC management
interfaces (e.g., OPC UA or embedded web interfaces), exploits the normal
overlap between loading and unloading operations by issuing small,
incremental adjustments to robot operating speed. These adjustments
slightly increase the VGR’s approach speed and gradually shrink the
temporal separation between the two robots.

\shortsectionBf{Detection.}
These manipulations remain within configured limits and do not
immediately trigger alarms. Once the attacker exceeds the maximum rate
limits on remote requests, the PLC detects the abnormal activity, blocks
further requests, raises an alarm, and reports the incident to the
operator.

\begin{figure}[t]
  \centering
  \includegraphics[width=.90\linewidth]{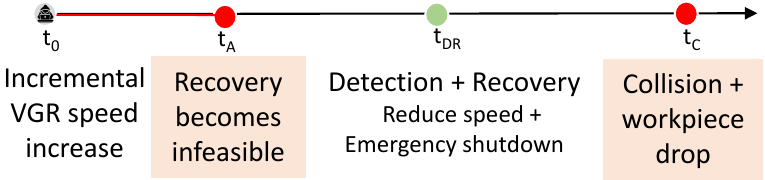}
  \caption{Successful detection but failed recovery leads to robot collision.
  }
  \label{fig:motivationAttack1}
\end{figure}

\shortsectionBf{Infeasible recovery.} Following detection, the operator activates the pre-defined recovery procedure that gradually reduces the robot speed. However, the attacker has already exhausted the remaining recovery margin. The available physical separation is insufficient for bounded deceleration to re-establish safe coordination before the robots enter the shared workspace, resulting in their collision. An emergency shutdown is not a viable option either, as abruptly stopping the robot while transporting a workpiece can compromise payload stability and cause the workpiece to be dropped, introducing a secondary safety hazard. Consequently, the deployed recovery procedure can no longer restore the system to a safe state without causing negative impact.

\subsection{Recovery-Unsafe Scenario: Toxic Gas Release}
\label{subsec:recovery-unsafe}
We illustrate a recovery-unsafe TLTR scenario in the chemical processing testbed. This example represents a broader class of industrial processes in which the available recovery actions remain executable after detection but, under attacker-induced process conditions, executing those actions introduces cascading safety hazards. Consequently, the recovery itself becomes unsafe even though the attack is correctly detected and recovery is triggered.


\shortsectionBf{Industrial process.} The chemical processing plant consists of an exothermic reactor whose byproducts are neutralized by a downstream scrubber. The reactor and scrubber are coordinated by the PLC and share cooling and neutralization resources.


\shortsectionBf{Recovery actions.} When anomalous reactor temperature or pressure is detected, the recovery actions will be to reduce reactant feed, increase coolant flow, and vent excess gases to the scrubber. These available recovery actions are intended to stabilize the reactor while maintaining safe downstream operation.


\shortsectionBf{Attack.} The attacker gradually increases reactant feed and slightly reduces coolant flow through legitimate PLC interfaces. These manipulations remain within the configured operating limits while stealthily increasing reactor temperature, pressure, and gas generation.


\shortsectionBf{Detection.} Once the attack is detected, the PLC raises an alarm, blocks attacker interactions, and transfers control to the operator to execute the available recovery actions.


\shortsectionBf{Unsafe recovery.} Upon detection, the operator activates the recovery actions by increasing cooling and venting gases to the scrubber. Although these actions initially stabilize the reactor, they also increase the demand on downstream neutralization resources. Because the attacker has already driven the process into a stressed operating condition before detection, continuing these recovery actions actually increases the risk of untreated toxic gas release. Consequently, the deployed recovery procedure can no longer be completed safely, rendering continued recovery unsafe despite correct detection. 


\begin{figure*}[t]
\centering
\includegraphics[width=.95\textwidth]{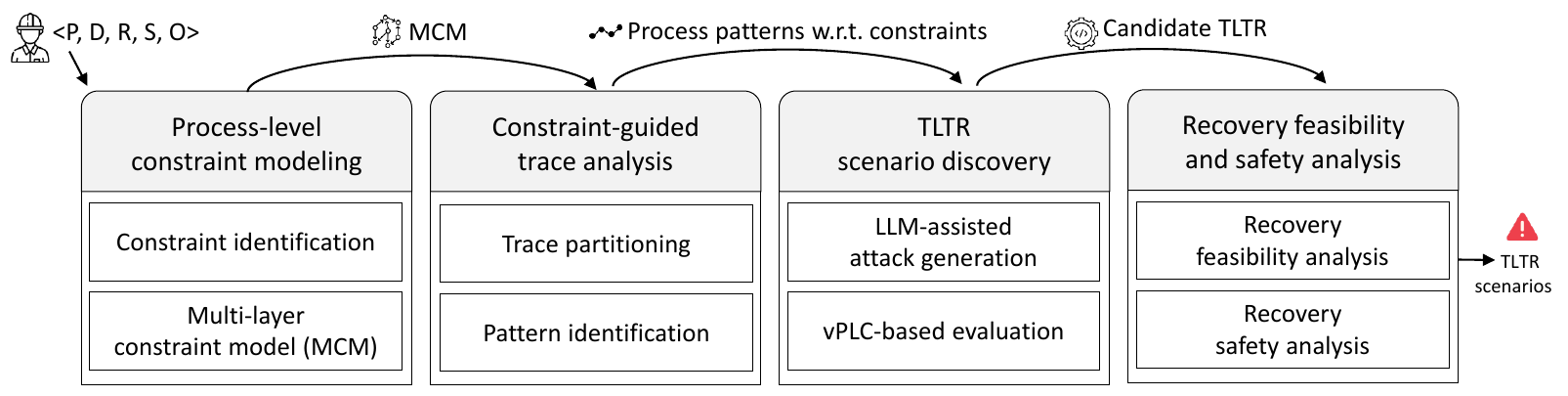}
\caption{Overview of \sys: The user provides P (PLC programs), D (detection policies), R (recovery policies), S (protection policies), and O (operational traces) of the subject ICS; \sys automatically discovers and validates TLTR scenarios.}   
\label{fig:overview}
\end{figure*}

\section{Design Goals and Challenges}
Our goal is to discover and confirm TLTR scenarios in which a subject ICS' recovery procedures become infeasible or unsafe after detection. This requires addressing three challenges.

\shortsectionBf{C1: Recoverability is not explicitly represented.} ICS design and operation typically prescribe when to detect abnormal behavior and how to respond, without knowing whether the deployed recovery procedure remains feasible and safe upon detection. This recoverability boundary is implicit in process dynamics, actuator limits, timing constraints, and interactions among control and safety mechanisms.

\shortsectionBf{C2: Detection does not guarantee recoverability.} An attack may be correctly detected before physical impact, yet the process state at detection may leave insufficient physical or temporal margin for the deployed recovery procedure to succeed. Identifying TLTR scenarios therefore requires reasoning about not only whether detection occurs, but whether sufficient recovery margin remains at that time. 

\shortsectionBf{C3: Recovery depends on evolving and interacting process conditions.} After detection, the ICS process state continues to evolve, and recovery actions may become ineffective or unsafe as margin continues to shrink or additional constraints are violated. Identifying TLTR scenarios therefore requires reasoning about how process state and constraint interactions evolve during recovery.

Together, these challenges motivate \sys's design for discovering and validating TLTR scenarios.
They also guide our evaluation, which examines effectiveness (RQ1, Section~\ref{subsec:rq1}), robustness (RQ4, Section~\ref{subsec:rq4}), and generality (RQ5, Section~\ref{subsec:rq5}). 

\section{\sys Overview}
\label{sec:overview}

\sys is an automated auditing/vetting framework for discovering \emph{too-late-to-recover (TLTR)} vulnerabilities in ICS. For a given ICS, the operator provides its PLC programs, detection and recovery policies, protection policies (where available), and operational traces (Figure~\ref{fig:overview}), based on which \sys automatically discovers and confirms TLTR scenarios with concrete attacks.


We illustrate \sys using a running example from the water treatment testbed, in which rising tank levels must be controlled through valve throttling to prevent overflow.

\shortsectionBf{(1) Process-level constraint modeling.}
\sys first builds a unified model of the target ICS by extracting constraints across the control, detection, and supervisory layers and organizing them into a multi-layer representation. A \emph{constraint} is represented as a conditional rule $(cond, act)$ over process variables, where $cond$ specifies the activation condition and $act$ specifies the enforced action.  Control constraints are derived from PLC programs using static analysis, while detection and recovery constraints are captured through structured rule parsing. These constraints define how control actions are applied, when detection is triggered, and what recovery actions are available. For example, in the water treatment system, a constraint such as $(\textsf{L1} \geq H_{1}^{\mathrm{alarm}}) \Rightarrow \textsf{Alarm}=1$ specifies the condition under which an alarm is triggered (Section~\ref{subsec:constraints}).

\shortsectionBf{(2) Constraint-guided trace analysis.}
Using the extracted constraints, \sys analyzes operational traces to characterize how the process evolves relative to these constraints during normal operation. In particular, it quantifies the remaining margin to relevant recovery-target or safety boundaries, how quickly this margin is consumed, and how effectively it can be restored through recovery actions. For example, in the water treatment system, \sys tracks how the tank level $\textsf{L1}$ approaches the alarm threshold over time and how quickly it can be reduced once corrective actions are applied (Section~\ref{subsec:pattern}). 

\shortsectionBf{(3) TLTR scenario discovery.} Using the multi-layer constraint model and the extracted constraint-centric patterns, \sys discovers candidate TLTR scenarios by exploring admissible process manipulations that reduce the recovery margin at detection. To navigate the combinatorial space of variables, timing, and manipulation magnitudes, \sys uses a constraint-guided large language model (LLM) agent to propose candidate attack scripts specifying which variables to manipulate, when to manipulate them, and by how much. The proposed manipulations are validated against constraint-derived bounds to ensure consistency with the attacker model and are then executed in a virtual PLC (vPLC) environment to obtain concrete process trajectories.


The space of admissible process manipulations is highly combinatorial due to interactions among multiple variables, layered constraints, and timing dependencies, making exhaustive rule-based enumeration impractical. Moreover, operational logs provide only partial coverage of system behavior, limiting purely data-driven exploration. In this setting, the LLM agent serves as a structured proposal mechanism for exploring combinations of variables, timing, and magnitudes under explicit constraint guidance. It does not perform feasibility reasoning or validation; all generated candidates are validated against the multi-layer constraint model and then executed in the vPLC environment to obtain their resulting process trajectories.


\textit{A TLTR scenario discovered by \sys.} 
In the water treatment system, the tank level is maintained at $\textsf{L1}=80$ during normal operation. The PLC applies corrective control when $\textsf{L1}$ enters the range $82$--$85$, the operator alarm is triggered when $\textsf{L1}$ reaches $90$, and the maximum capacity of the tank is $\textsf{L1}=100$. \sys discovers a TLTR scenario in which the process is manipulated prior to the alarm by progressively increasing inflow and reducing drainage. When $\textsf{L1}$ enters the pre-alarm range $[88,90)$, the attacker maximizes inflow and suppresses drainage, causing the level to rise rapidly. Because alarm evaluation and supervisory response occur at discrete control cycles rather than continuously, the level continues increasing before recovery actions take effect. As a result, when recovery is initiated, the tank level has already reached $\textsf{L1}=96$. The operator then applies the correct recovery action by closing the inlet valve and enabling drainage. However, closing the inlet valve takes approximately $3$ seconds. During this transient, the level continues increasing at approximately $1.5$ units/s and exceeds the maximum tank capacity before recovery becomes effective. Thus, the tank overflows despite correct detection and response (Section~\ref{subsec:incident_generation}). 

{\em The attacker does not need to remain active until overflow ($\textsf{L1} > 100$); it only needs to drive the system to a state where the remaining recovery margin is insufficient once recovery begins ($\textsf{L1}=96$).}

\shortsectionBf{(4) Recovery feasibility and safety analysis.}
Finally, \sys analyzes each executed scenario to determine whether recovery remains feasible once detection occurs. It examines how the process evolves after the alarm is triggered and whether the deployed recovery logic can restore the process to a safe operating state without violating applicable constraints. In the water treatment example, once the alarm is raised, recovery actions (e.g., reducing inflow and enabling drainage) are applied under system limits. However, because the remaining margin at detection is already small, these actions cannot reduce the tank level $\textsf{L1}$ quickly enough to prevent violation. Based on this analysis, \sys classifies each TLTR scenario as \emph{recovery-infeasible} if the deployed recovery procedure cannot reach its recovery condition within the specified recovery interval or cannot prevent violation of its target constraint, or \emph{recovery-unsafe} if execution of the recovery procedure is associated with a secondary constraint violation before the recovery condition is reached (Section~\ref{sec:recovery_analysis}).

\section{\sys Design}
\label{sec:methodology}
\shortsectionBf{Recoverability and TLTR condition.}
Let $x_d$ denote the process state when detection occurs and $\rho$ the deployed recovery logic, including its specified recovery interval. We consider $x_d$ \emph{recoverable} if, starting from $x_d$, executing $\rho$ can reach its specified recovery condition within this interval without a target-constraint violation or a recovery-associated secondary-constraint violation. Conversely, $x_d$ is \emph{too late to recover} if $\rho$ cannot reach this recovery condition within the interval without such a violation. Accordingly, an ICS exhibits a TLTR vulnerability if there exists an admissible attack scenario that drives the process to a too-late-to-recover state at detection while the deployed detection and recovery mechanisms remain operational as intended. The recovery condition is defined by the completion conditions of the deployed recovery logic and represents the operating state that $\rho$ is intended to restore. \sys evaluates recoverability with respect to the deployed recovery logic $\rho$, rather than all physically conceivable recovery or protection strategies. Accordingly, activation of a higher-level protection mechanism such as SIS or ESD does not constitute successful completion of $\rho$; instead, it indicates that the deployed recovery procedure could no longer restore its intended recovery condition and the process had to escalate to fail-safe protection.


\subsection{Process-level Constraint Modeling}
\label{subsec:constraints}

\shortsectionBf{Constraint identification.} \sys extracts constraints from PLC programs and machine-parsable detection, recovery, and protection policies. \sys analyzes PLC programs (IEC 61131-3 ST/SCL) using static analysis over control-flow and data-dependence graphs to identify condition--action logic governing process variables, including actuator limits, rate constraints, and interlocks. Detection, recovery, and, where available, protection policies (e.g., operator emergency shutdown procedures, ESD, or SIS policies) are parsed as ``condition--action'' specifications to extract alarm-triggering conditions, recovery actions, and their specified completion conditions.


Each extracted constraint $C_i$ is represented as a conditional rule $(cond_i, act_i)$, where $cond_i$ is a Boolean predicate over process variables, and $act_i$ specifies the corresponding enforcement action. Each condition $cond_i$ may involve multiple process variables, allowing \sys to represent constraints whose activation depends on combined process conditions. 

\shortsectionBf{Multi-layer constraint model.}
\sys constructs a multi-layer constraint model (MCM) from the extracted constraints. The MCM is a directed graph $G = (C, E)$, where each node $C_i \in C$ represents a constraint annotated with its triggering condition, enforcement action, and regulated variables. When $cond_i$ holds, $act_i$ specifies the corresponding control,
detection, or recovery action.

Edges $(C_i, C_j) \in E$ capture relationships between constraints, derived from (i) implication between activation conditions (e.g., $cond_i \Rightarrow cond_j$), (ii) shared regulated variables, and (iii) interactions across control, detection, and supervisory logic. An edge indicates a dependency or interaction between two constraints
as process conditions evolve. The quantitative characteristics of these relationships are derived in the subsequent analysis (Section~\ref{subsec:pattern}).

\begin{figure}[t]
  \centering
  \includegraphics[width=.95\linewidth]{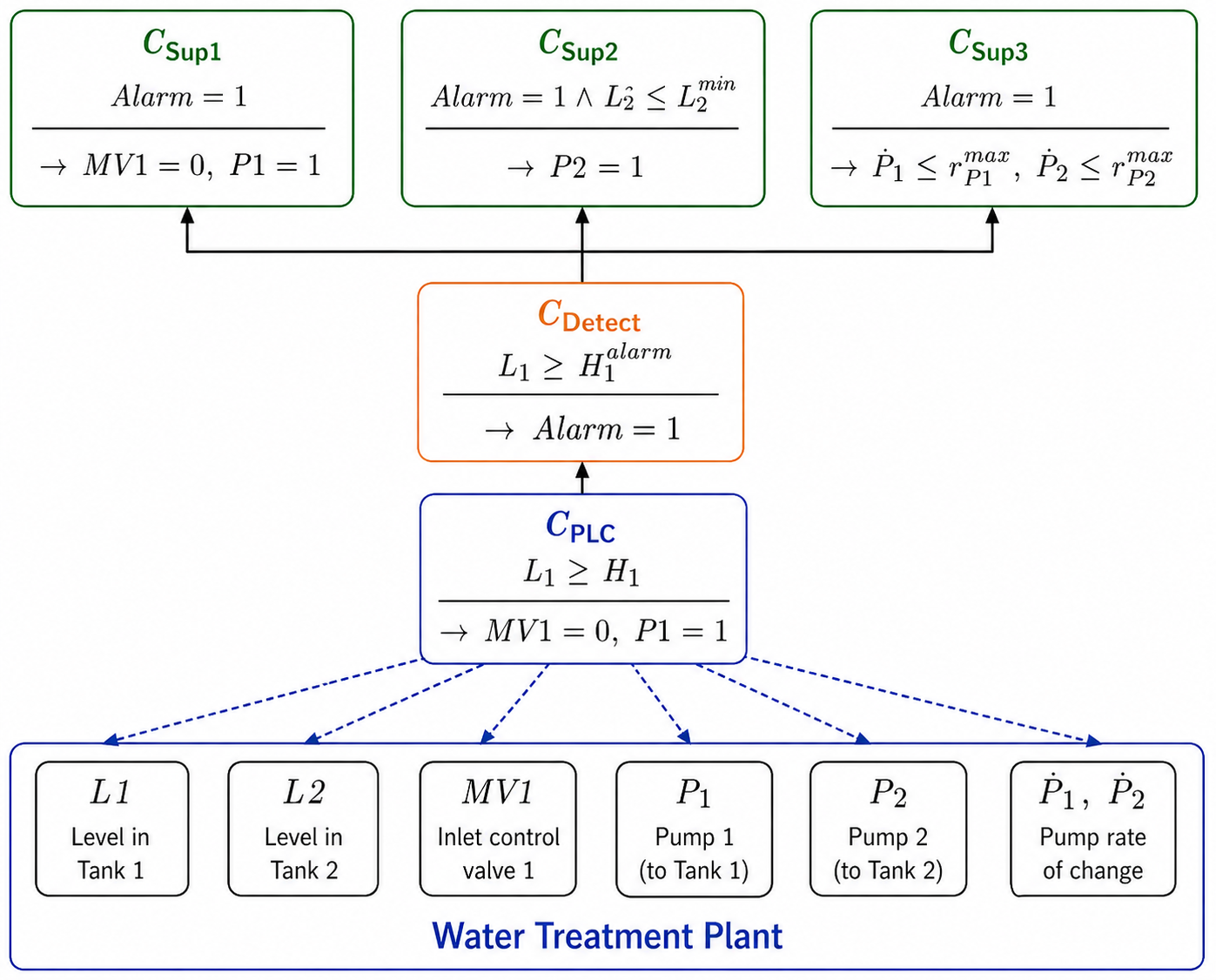}
  \caption{Multi-layer constraint model (MCM) for the water treatment system.
  }
  \label{fig:mcm}
\end{figure}

\sys builds the MCM by grouping constraints that regulate overlapping variables and linking them according to the relationships defined above. This construction captures how PLC control, detection, and supervisory recovery constraints interact as process conditions evolve, while simultaneously restricting or modifying the set of admissible actions. The following constraints define the MCM for the water treatment system shown in Figure~\ref{fig:mcm}.

\begin{tcolorbox}[colback=white!0,colframe=gray!0,
  boxsep=1pt,
  top=.1pt,
  bottom=.1pt,
  left=2pt,
  right=2pt]
\noindent
$C_{\mathrm{PLC}}$: $(\textsf{L1} \geq H_1) \Rightarrow (\textsf{MV1}=0 \wedge \textsf{P1}=1)$.\

$C_{\mathrm{Detect}}$: $(\textsf{L1} \geq H_{1}^{\mathrm{alarm}}) \Rightarrow \textsf{Alarm}=1$.\

$C_{\mathrm{Supervisory},1}$: $(\textsf{Alarm}=1) \Rightarrow (\textsf{MV1}=0 \wedge \textsf{P1}=1)$.\

$C_{\mathrm{Supervisory},2}$: $(\textsf{Alarm}=1 \wedge \textsf{L2} \leq L_{2}^{\min}) \Rightarrow \textsf{P2}=1$.\

$C_{\mathrm{Supervisory},3}$: $(\textsf{Alarm}=1) \Rightarrow (\dot{\textsf{P1}} \leq r_{\textsf{P1}}^{\max} \wedge \dot{\textsf{P2}} \leq r_{\textsf{P2}}^{\max})$.
\end{tcolorbox}

The PLC constraint $C_{\mathrm{PLC}}$ enforces coordinated actuation by shutting down inflow and activating drainage when the upstream tank level exceeds $H_1$, defining a restricted actuation region prior to detection. The detection constraint $C_{\mathrm{Detect}}$ raises an alarm at a higher threshold $H_{1}^{\mathrm{alarm}}$, marking the transition from automated control to supervisory intervention. The supervisory constraints define the recovery behavior after detection: $C_{\mathrm{Supervisory},1}$ preserves the shutdown-and-drain response, $C_{\mathrm{Supervisory},2}$ enables downstream transfer when the second-stage level approaches its lower bound, and $C_{\mathrm{Supervisory},3}$ bounds actuator adjustment rates during recovery. 

\subsection{Constraint-Guided Trace Analysis}
\label{subsec:pattern}
While the MCM (Section~\ref{subsec:constraints}) captures the constraints that govern process behavior, this step analyzes how the process evolves relative to these constraints under normal operation. In particular, \sys quantifies how recovery margin is consumed, how quickly constraint boundaries are approached, and how effectively recovery actions restore that margin.

\begin{figure}[t]
  \centering
  \includegraphics[width=.95\linewidth]{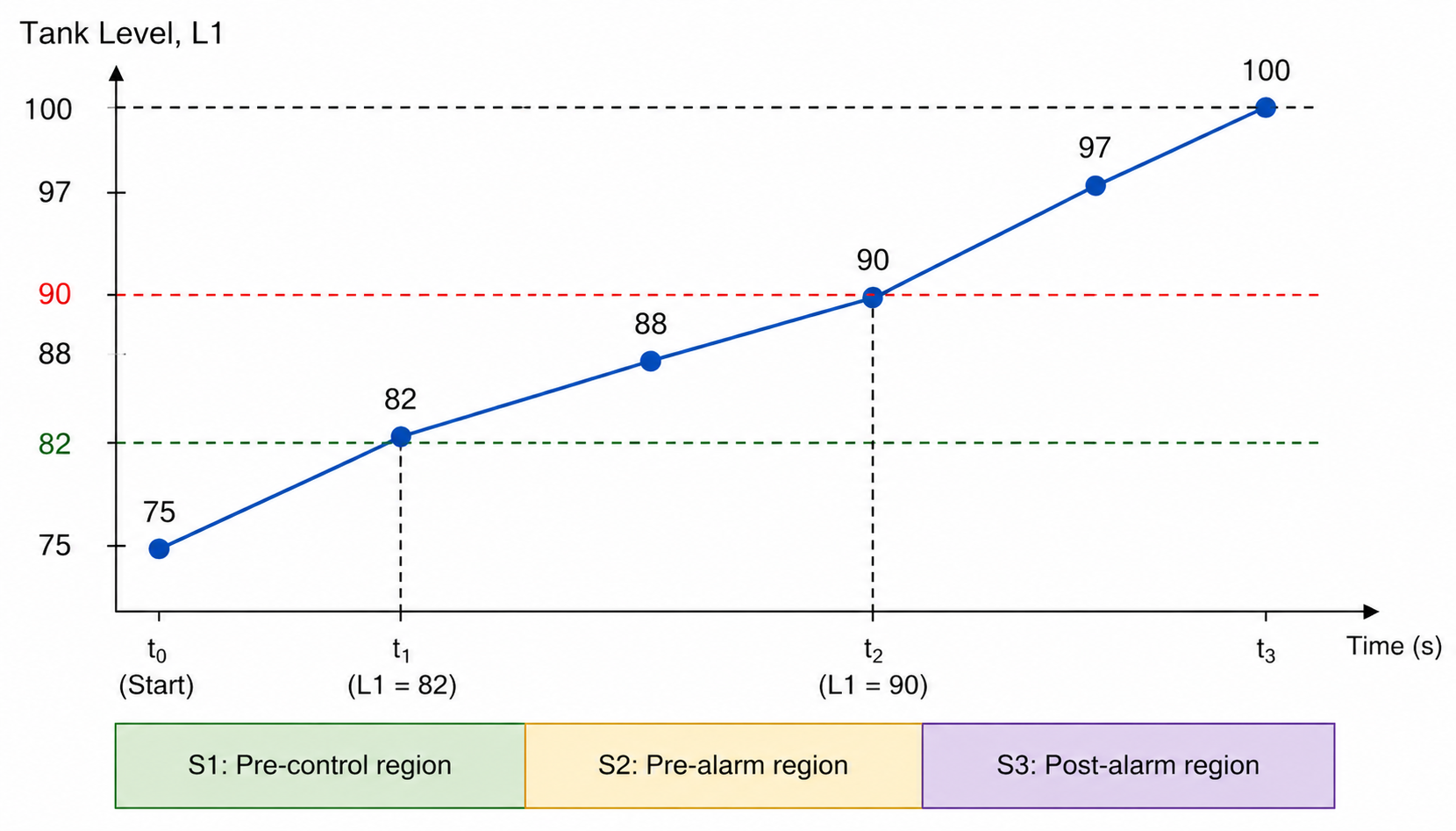}
  \caption{Trace partitioning based on constraint activation.
  }
  \label{fig:trace_partitioning}
\end{figure}

\shortsectionBf{Trace partitioning.} \sys partitions operational traces into constraint-aligned segments $S_i$ by identifying points at which relevant constraints become active or inactive. Each segment corresponds to a region of the trace with a distinct constraint-activation pattern. For each constraint of interest, \sys associates the corresponding trace region with that constraint for subsequent margin analysis. If a constraint is not explicitly activated, \sys uses the time at which its slack is minimized as a proxy for the closest approach to its boundary. Figure~\ref{fig:trace_partitioning} illustrates this for the water treatment plant. The trace is split at the PLC control threshold ($L1=82$) and the alarm threshold ($L1=90$), generating three segments.


\shortsectionBf{Pattern identification.}
For each segment $S_i$, \sys analyzes the operational trace and computes a slack function with respect to the relevant recovery-target or safety boundary identified by the MCM. Using the slack values over the segment, \sys derives three quantities:
\[
\begin{aligned}
D_i &= \min_{t\in S_i} slack_i(t), \qquad
R_i = \max_{t\in S_i}\dot{slack}_i(t),\\
T_i &= \min_{\substack{t\in S_i\\ \dot{slack}_i(t)<0}}
\frac{slack_i(t)}{|\dot{slack}_i(t)|}.
\end{aligned}
\]
where $D_i$ is the minimum remaining slack to the relevant recovery-target, $T_i$ is the minimum estimated time to reach that boundary while the process is approaching it, and $R_i$ is the maximum observed rate at which slack is restored within the segment.

Each segment is summarized as a numeric pattern set over $(D_i, T_i, R_i)$. Together, these quantities characterize the available recovery margin: $D_i$ captures the remaining physical slack, $T_i$ captures how quickly that slack may be exhausted, and $R_i$ captures the observed ability of recovery actions to restore it. In addition, \sys records the corresponding process state at the time where the minimum slack occurs, providing a reference point for each pattern.

\noindent\textbf{Example.}
Consider segments from normal operation traces in Figure~\ref{fig:trace_partitioning}. In segment $S_1$ (pre-control region), the trace is:
\[
60,\,65,\,67,\,70
\]
At this stage, the process remains far from the relevant safety boundary, yielding large values of $D_1$ and $T_1$ and therefore substantial remaining recovery margin.

In segment $S_2$ (pre-alarm region), the trace evolves as:
\[
82,\,85,\,88,\,90.
\]
Here, the system approaches the alarm threshold, leaving a physical slack of $100 - 90 = 10$ units to the safety boundary.

Following detection, in segment $S_3$ (post-alarm region), recovery actions are initiated. Due to actuation delays, the level continues to rise briefly before decreasing:
\[
90,\,92,\,94,\,95,\,94,\,92,\,88.
\]
From this, \sys extracts the recovery pattern:
\[
P_{S_3} = \{D_3 \approx 5,\; T_3 \approx 6~\text{s},\; R_3 \approx 2\}.
\]
This pattern shows that, under normal operation, the level peaks below the safety limit and recovery actions restore the system. However, the small remaining recovery margin at detection indicates that modest deviations could move the process to a state where the recovery actions are no longer sufficient.

\subsection{TLTR Scenario Discovery}
\label{subsec:incident_generation}



\sys discovers candidate TLTR scenarios by exploring admissible process manipulations that reduce the recovery margin at detection relative to the constraint-centric patterns identified above. In particular, the extracted values of $D_i$, $T_i$, and $R_i$ identify constraint regions in which little physical or temporal margin remains and recovery is comparatively limited. \sys uses these low-margin regions, together with the MCM, to guide an LLM agent in proposing \emph{candidate attack scripts}, i.e., structured specifications of process-variable manipulations intended to drive the process toward such regions before detection.

An attack script $\alpha$ is a time-ordered sequence of condition--action specifications together with an observation horizon $H$. Each specification defines (i) a trigger condition over process variables and (ii) a parameterized manipulation, including the variables modified, manipulation magnitude, and duration. During execution, each manipulation is applied when its corresponding trigger condition becomes true.

The observation horizon $H$ defines a post-detection interval during which no further adversarial manipulations are applied, allowing recoverability to be evaluated under the deployed control and recovery logic. The LLM proposes scripts with approximate parameters, which \sys validates against constraint-derived bounds to ensure that the manipulations are admissible under the attacker model. The validated scripts are then executed in the vPLC to obtain concrete process trajectories for subsequent recovery analysis.

\shortsectionBf{LLM-assisted attack generation.}
\sys encodes the multi-layer constraint model and constraint-centric patterns into structured prompts that specify controllable variables, admissible ranges, and process dynamics. To target different recovery failure modes, \sys uses goal-oriented prompts for recovery-infeasible and recovery-unsafe scenarios (Appendix~\ref{app:llm_prompts}).

Guided by these inputs, the LLM agent generates candidate attack scripts by selecting trigger conditions, variables, and approximate timing and magnitude. For example, a multi-stage attack script ($\alpha_{\mathrm{WP}}$) generated by \sys for the water treatment system is shown below:

\[
\alpha_{\mathrm{WP}} =
\begin{cases}

\textbf{Stage 1 (Pre-conditioning):} \\
\quad \text{if } L1 \in [80,84]: \\
\quad\quad P1 \leftarrow 0.7 \cdot P1_{\max} \\

\textbf{Stage 2 (Margin compression):} \\
\quad \text{if } L1 \in [84,88]: \\
\quad\quad MV1 \leftarrow 0.8 \cdot MV1_{\max}, \quad
P1 \leftarrow 0.5 \cdot P1_{\max} \\

\textbf{Stage 3 (Pre-alarm push):} \\
\quad \text{if } L1 \in [88,90): \\
\quad\quad MV1 \leftarrow MV1_{\max}, \quad P1 \leftarrow 0 \\

\textbf{Stage 4 (Post-detection):} \\
\quad \text{if } Alarm = 1: \\
\quad\quad \text{stop attack and observe for } H
\end{cases}
\]

In this attack, the process initially operates near its nominal setpoint ($L1 \approx 80$). Stage~1 slightly reduces drainage, increasing net inflow without triggering alarms. Stage~2 gradually increases inflow while further limiting drainage, causing the level to rise slowly (e.g., $82 \rightarrow 85 \rightarrow 88$) while remaining below the alarm threshold ($L1 = 90$). Stage~3 is triggered in the pre-alarm range $L1 \in [88,90)$, where inflow is maximized and drainage is suppressed. Since alarm evaluation and supervisory response occur at discrete control cycles, the level can continue rising after the stage is triggered and before recovery actions take effect. Thus, by the time recovery begins, $L1$ has reached a low-margin state (e.g., $L1 \approx 96$), leaving insufficient margin for the deployed recovery logic to prevent overflow under bounded valve actuation.


\sys uses GPT-4~\cite{@gpt4} via the OpenAI API~\cite{@openai_api}, but the approach is model-agnostic.

\shortsectionBf{vPLC-based evaluation.}
For each attack script $\alpha$, \sys validates parameters against constraint-derived bounds and executes the script in a sandboxed vPLC environment that reproduces PLC logic, scheduling, actuator limits, and alarms. Each execution produces a time-stamped trace of system behavior, including variables, actuator commands, and alarm states. These traces are analyzed to determine whether recovery remains feasible or becomes infeasible or unsafe.

\begin{algorithm}[t]
    \small
    \caption{Recovery feasibility and safety analysis}
    \label{alg:recovery}
    \begin{algorithmic}[1]
        \Require{Execution trace $T$, multi-layer constraint model $\mathcal{M}$}
        \Ensure{Classification $\in$ \{\emph{Recoverable}, \emph{Infeasible}, \emph{Unsafe}\}}

        \State $t_d \gets \text{GetDetectionTime}(T)$
        \State $t_c \gets \text{GetRecoveryCompletionTime}(T,\mathcal{M},t_d)$

        \State $\mathcal{V} \gets \text{GetConstraintViolations}(T)$
        \State $\mathcal{V} \gets \text{SortByTime}(\mathcal{V})$

        \State $\mathcal{C}_{\rho} \gets
        \text{GetRecoveryTargetConstraints}(\mathcal{M})$

        \State $\mathcal{C}_{sec} \gets
        \text{GetRecoveryAffectedConstraints}(\mathcal{M})
        \setminus \mathcal{C}_{\rho}$

        \For{each $(C_i,t_i) \in \mathcal{V}$}

            \If{$C_i \in \mathcal{C}_{\rho}$ \textbf{and}
                $t_i \geq t_d$ \textbf{and}
                ($t_c = \emptyset$ \textbf{or} $t_i < t_c$)}
                \State \Return Infeasible
            \EndIf

            \If{$C_i \in \mathcal{C}_{sec}$ \textbf{and}
                $\text{RecoveryAssociated}(C_i,T)$ \textbf{and}
                ($t_c = \emptyset$ \textbf{or} $t_i < t_c$)}
                \State \Return Unsafe
            \EndIf

        \EndFor

        \If{$t_c = \emptyset$}
            \State \Return Infeasible
        \EndIf

        \State \Return Recoverable

    \end{algorithmic}
\end{algorithm}

\subsection{Recovery Feasibility and Safety Analysis}
\label{sec:recovery_analysis}
\sys analyzes each scenario execution trace to determine whether the deployed recovery logic can successfully and safely restore the process after detection. For each deployed recovery procedure, \sys distinguishes the \emph{target constraint} that the procedure is intended to restore from \emph{secondary constraints} that may be affected during recovery. Recovery is \emph{infeasible} if the deployed recovery logic either fails to reach its specified recovery condition within the recovery interval or a target constraint is violated before that condition is reached, and \emph{unsafe} if execution of the deployed recovery logic is associated with a violation of a secondary constraint before the recovery condition is reached.



\sys performs this analysis using Algorithm~\ref{alg:recovery}. Let $t_d$ denote the detection time and $t_c$ the first timestamp at or after $t_d$ at which the specified recovery completion conditions are satisfied. These completion conditions define the safe state that the deployed recovery logic is intended to reach. As the process evolves during recovery, additional recovery actions may be triggered according to their specified conditions and are included in the execution of $\rho$. The observation horizon $H$ is selected to cover the recovery interval specified by the recovery logic. \sys classifies the execution according to the first constraint violation after detection and before recovery completion: a target-constraint violation yields \emph{recovery-infeasible}, whereas a recovery-associated secondary-constraint violation yields \emph{recovery-unsafe}. If no such violation occurs but the recovery condition is not reached within the recovery interval, the execution is classified as \emph{recovery-infeasible}.

Next, \sys extracts all constraint violations from the trace and orders them chronologically. Using the multi-layer constraint model, \sys identifies the target constraints that the recovery procedure is intended to restore and the secondary constraints that may be affected by its recovery actions.

If a target constraint is violated after detection and before recovery completes, \sys classifies the execution as \emph{recovery-infeasible}, indicating that the deployed recovery logic cannot restore its target condition before the corresponding violation occurs. A secondary violation is considered \emph{recovery-associated} if (i) it occurs after the corresponding recovery action is initiated, (ii) the violated variable is affected by that action through a dependency represented in the MCM, and (iii) the observed response following the recovery action moves the affected variable toward the violated boundary. Accordingly, $\text{RecoveryAssociated}(C_i,T)$ evaluates these conditions using the execution trace to identify the initiation time of the corresponding recovery action and the dependencies encoded in the MCM. If such a secondary violation occurs before the recovery condition is reached, \sys classifies the execution as \emph{recovery-unsafe}. If the recovery condition is reached without either a target-constraint violation or a recovery-associated secondary-constraint violation, the execution is labeled \emph{Recoverable}.

\begin{table}[t!]
\centering
\footnotesize
\caption{Industrial testbeds used in evaluation.}
\label{tab:testbeds}
\renewcommand{\arraystretch}{1.1}
\setlength{\tabcolsep}{6pt}
\begin{threeparttable}
\begin{tabular}{|c|c|c|}
\hline
\textbf{Testbed} & \textbf{Physical PLC} & \textbf{Virtual PLC} \\
\hline
MP 
& Siemens S7-1500 (CPU 1512SP) 
& SIMATIC S7-1500V \\ 
\hline
CP 
& PLCnext Control AXC F 2152 
& Control 2000 (vPLCnext) \\ 
\hline
WP 
&  CODESYS Control  (Pi 4)
& CODESYS Control (SL) \\ 
\hline
\end{tabular}

\begin{tablenotes}[flushleft]
\footnotesize
\item \textbf{MP}: Manufacturing plant, \textbf{CP}: Chemical plant, \textbf{WP}: Water treatment plant.
\end{tablenotes}
\end{threeparttable}
\end{table}

\section{Evaluation}
\label{sec:eval}

\subsection{Experimental Setup}
\label{subsec:exp_setup}


We evaluate \sys on three industrial testbeds that capture diverse control environments and recovery mechanisms: a Fischertechnik manufacturing plant (MP) ~\cite{@fp} controlled by a Siemens S7-1500~\cite{SiemensFirmwareupdat}, a chemical processing system (CP) ~\cite{@cp} controlled by a PLCNext~\cite{@storePLCNext}, and a water treatment plant (WP)~\cite{@wp} controlled by a CodeSys PLC~\cite{@codesys}. Table~\ref{tab:testbeds} summarizes the corresponding physical and virtual PLCs used across these platforms. Fischertechnik plant is a miniature factory testbed, whereas the chemical plant and water treatment system are hybrid platforms with simulated physical processes controlled by real PLCs. These platforms implement realistic control logic and are widely used as benchmarks in prior ICS security studies~\cite{@SAIN,vetplc,adepu2016generalized}.

We use a virtual PLC (vPLC) environment for large-scale exploration and
preliminary screening of candidate attack scripts, enabling safe and
efficient evaluation. Validated trajectories are then executed on
physical and high-fidelity testbeds with real PLCs in the loop to
confirm that the identified recovery failures persist under realistic
timing, actuation, and safety constraints.

\textit{ICS Settings.}
We evaluate each testbed under two ICS settings. The {\em original} setting uses the original control logic, alarm mechanisms, and available operator actions (e.g., HMI commands and observed control behaviors) as implemented in the testbeds, serving as the baseline (Table~\ref{tab:rq1_original}). These settings reflect typical operational deployments and are not explicitly designed to ensure recovery feasibility under adversarial conditions. For these testbeds, we derive the recovery procedures from existing PLC control logic, HMI-exposed operator actions, and operator-driven control patterns observed in operational traces, without introducing new recovery capabilities. The evaluated testbeds do not include independently engineered SIS/ESD systems. Accordingly, this evaluation focuses on post-detection recovery. When protection policies are available, \sys can incorporate them into its analysis, as demonstrated in the industrial fertilizer plant case study (Section~\ref{sec:industry}).

To assess whether improved detection and recovery capabilities reduce TLTR exposure, we construct a {\em security-enhanced} configuration for each testbed. The enhancements are selected independently of the TLTR scenarios discovered by RISK and are derived only from the corresponding system specifications, admissible operating limits, hazard-analysis principles, and prior safety studies~\cite{castellanos2023provable, khalid2025assessing}. Specifically, we consider standard safety-engineering measures such as earlier alarm triggering within admissible operating ranges, rate-limited actuation, and additional graduated recovery actions permitted by the system. The resulting configuration is fixed before re-evaluating the RISK-generated scenarios and is applied uniformly rather than customized to individual attacks. Thus, the enhanced setting represents a generally strengthened detection--recovery configuration, rather than a collection of attack-specific mitigations. RQ1 (Section~\ref{subsec:rq1}) compares the original and enhanced configurations to quantify how improved recovery capability affects TLTR outcomes; all remaining experiments use the original configuration to maintain a consistent basis for comparison across methods, defenses, and system variations.


During evaluation, an automated recovery agent executes the available recovery actions according to their predefined triggering conditions. The agent continuously monitors process variables through standard human--machine interfaces (HMI) and applies corrective actions. This setup reflects supervisory-layer automation commonly used in ICS, where predefined control responses and operator assistance workflows are implemented through HMI/SCADA systems~\cite{mohammad2025scada}. 




\shortsectionBf{Research Questions.}
We present our evaluation results that answer the following research questions:


\begin{enumerate}[leftmargin=9mm,topsep=.5mm] %
\setlength{\itemsep}{-0.2mm}
\item[\textbf{RQ1}] How effectively does \sys identify TLTR scenarios across the evaluated ICS?
\item[\textbf{RQ2}]
How does \sys compare with existing ICS attack generation methods?
\item[\textbf{RQ3}]
How effective are state-of-the-art ICS defenses against \sys-generated TLTR attacks?
\item[\textbf{RQ4}] How sensitive is \sys to variations in traces and LLM-generated candidate attack scripts?
\item[\textbf{RQ5}] How well can \sys generalize across different ICS platforms and industrial processes?


\end{enumerate}


\begin{table}[t!]
\centering
\small
\caption{TLTR outcomes under original ICS configurations.}
\label{tab:rq1_original}
\setlength{\tabcolsep}{4pt}
\renewcommand{\arraystretch}{1.08}

\begin{threeparttable}
\begin{tabular}{|c|c|c|c|c|}
\hline
\textbf{Testbed} &
\textbf{Evaluated} &
\textbf{TLTR} &
\textbf{Inf.} &
\textbf{Uns.} \\ \hline\hline

MP & 188 & 137 (73\%) & 79 & 58 \\ \hline
CP & 171 & 132 (77\%) & 73 & 59 \\ \hline
WP & 158 & 123 (78\%) & 70 & 53 \\ \hline

\textbf{Total} &
\textbf{517} &
\textbf{392 (76\%)} &
\textbf{222} &
\textbf{170} \\ \hline

\end{tabular}

\begin{tablenotes}[flushleft]
\scriptsize
\item Inf.: recovery-infeasible; Uns.: recovery-unsafe.
\end{tablenotes}

\end{threeparttable}
\end{table}

\begin{table}[t!]
\centering
\small
\caption{Effect of security-enhanced configurations.}
\label{tab:rq1_enhanced}
\setlength{\tabcolsep}{4pt}
\renewcommand{\arraystretch}{1.08}

\begin{threeparttable}
\begin{tabular}{|c|c|c|c|}
\hline
\textbf{Testbed} &
\textbf{Original TLTR} &
\textbf{Enhanced TLTR} &
\textbf{Mitigated} \\ \hline\hline

MP & 137 (73\%) & 38 (20\%) & 99 (72\%) \\ \hline
CP & 132 (77\%) & 39 (23\%) & 93 (70\%) \\ \hline
WP & 123 (78\%) & 34 (22\%) & 89 (72\%) \\ \hline

\textbf{Total} &
\textbf{392 (76\%)} &
\textbf{111 (21\%)} &
\textbf{281 (72\%)} \\ \hline

\end{tabular}

\begin{tablenotes}[flushleft]
\scriptsize
\item Mitigated percentage is relative to the original TLTR scenarios.
\end{tablenotes}

\end{threeparttable}
\end{table}

\subsection{\sys Effectiveness (RQ1)}
\label{subsec:rq1} We evaluate how effectively \sys identifies TLTR scenarios across the three testbeds. For each testbed, \sys generates candidate attack scripts using the multi-layer constraint model and constraint-centric patterns extracted from operational traces. Each candidate is validated against the attacker model and constraint-derived manipulation bounds, executed in the corresponding vPLC environment, and analyzed using Algorithm~\ref{alg:recovery}. The resulting executions are classified as \emph{Recoverable}, \emph{Recovery-Infeasible}, or \emph{Recovery-Unsafe}. We first evaluate the original ICS configurations and then re-evaluate the same scenarios under the security-enhanced configurations.

\shortsectionBf{TLTR scenarios under the original configurations.} Table~\ref{tab:rq1_original} summarizes the outcomes under the original ICS configurations. Across the three testbeds, \sys evaluates 517 candidate scenarios and identifies 392 TLTR outcomes (76\%). Specifically, 137 of 188 scenarios (73\%) in MP, 132 of 171 (77\%) in CP, and 123 of 158 (78\%) in WP result in a recovery-infeasible or recovery-unsafe condition at detection. The remaining 125 scenarios are recoverable under the deployed recovery logic.

\shortsectionBf{Recovery failure modes.}
Among the 392 TLTR scenarios, 222 (57\%) are recovery-infeasible and 170 (43\%) are recovery-unsafe. Recovery-infeasible scenarios occur when the deployed recovery logic cannot reach its specified recovery condition within the available physical or temporal constraints, or when a target constraint is violated before recovery completes. Recovery-unsafe scenarios occur when execution of the deployed recovery logic is associated with a secondary constraint violation before the recovery condition is reached. Appendix~\ref{app:tltr_examples} reports ten TLTR scenarios from MP, CP, and WP, including their pre-detection manipulations, states at detection, deployed recovery actions, and post-detection outcomes.

\shortsectionBf{Insufficient recovery margin at detection.} We further quantify the temporal recovery margin available when detection occurs. Table~\ref{tab:margin_deficit} reports the minimum, median, and maximum insufficient recovery margin observed across the three testbeds. The median margin is 5\,s in MP, 18\,s in CP, and 7\,s in WP, with observed values ranging from  1\,s to 45\,s. These results show that, for the identified TLTR scenarios, detection can occur after the available recovery margin has already become insufficient for the deployed recovery procedure to complete.

\begin{table}[t!]
\centering
\small
\caption{Insufficient recovery margin at detection (seconds).}
\label{tab:margin_deficit}
\setlength{\tabcolsep}{8pt}
\renewcommand{\arraystretch}{1.1}
\begin{tabular}{|c|c|c|c|}
\hline
\textbf{Testbed} & \textbf{Min} & \textbf{Median} & \textbf{Max} \\ \hline\hline
MP & 1 & 5 & 12 \\ \hline
CP & 1 & 18 & 45 \\ \hline
WP & 4 & 7 & 16 \\ \hline
\end{tabular}
\end{table}

\shortsectionBf{Effect of security-enhanced configurations.} We next re-evaluate the same 517 scenarios under the security-enhanced configurations. As shown in Table~\ref{tab:rq1_enhanced}, TLTR outcomes decrease from 392 (76\%) to 111 (21\%). In MP, the number decreases from 137 to 38; in CP, from 132 to 39; and in WP, from 123 to 34. Overall, 281 of the 392 TLTR scenarios identified under the original configurations no longer result in TLTR outcomes under the evaluated enhancements, corresponding to a 72\% reduction. The remaining 111 scenarios continue to reach recovery-infeasible or recovery-unsafe states under the strengthened configurations.

\begin{table}[t!]
\centering
\small
\caption{Ablation of \sys's TLTR scenario discovery.}
\label{tab:ablation}
\setlength{\tabcolsep}{4pt}
\renewcommand{\arraystretch}{1.08}

\begin{tabular}{|c|c|c|c|c|}
\hline
\textbf{Configuration} &
\textbf{MP} &
\textbf{CP} &
\textbf{WP} &
\textbf{Overall} \\ \hline\hline

Full \sys & 73\% & 77\% & 78\% & \textbf{76\%} \\ \hline
w/o MCM & 37\% & 43\% & 39\% & \textbf{40\%} \\ \hline
w/o patterns & 49\% & 51\% & 53\% & \textbf{51\%} \\ \hline
TLTR objective only & 12\% & 17\% & 13\% & \textbf{14\%} \\ \hline

\end{tabular}
\end{table}

\shortsectionBf{Ablation study.}
We evaluate the contribution of the two inputs used to guide \sys's TLTR scenario discovery: the multi-layer constraint model (MCM) and constraint-centric patterns. We separately remove each input and then retain only the TLTR objective, while keeping the remaining validation and execution procedure unchanged. As shown in Table~\ref{tab:ablation}, removing the MCM reduces the overall TLTR rate from 76\% to 40\%, while removing the constraint-centric patterns reduces it to 51\%. When only the TLTR objective is retained, the TLTR rate decreases to 14\%. These results show that both the MCM and constraint-centric patterns substantially contribute to \sys's ability to discover TLTR scenarios.

\begin{tcolorbox}[colback=gray!8,colframe=gray!40,boxrule=0.5pt]
These results further demonstrate that successful attack detection does not necessarily imply successful recovery. Even under security-enhanced configurations, some attacks can still drain the available recovery margin before detection, such that the deployed recovery procedure cannot restore the process to its intended safe operating state.
\end{tcolorbox}

\begin{table}[t!]
\centering
\footnotesize
\caption{\sys vs existing attack-generation methods.}
\label{tab:rq2_compare}
\setlength{\tabcolsep}{4pt}
\renewcommand{\arraystretch}{1.08}

\begin{tabular}{|c|c|c|c|}
\hline
\textbf{Testbed} &
\textbf{Method} &
\textbf{Evaluated} &
\textbf{TLTR} \\ \hline\hline

\multirow{3}{*}{MP}
& AttackLLM & 180 & 12 (7\%) \\ \cline{2-4}
& GENICS    & 185 & 14 (8\%) \\ \cline{2-4}
& \sys      & 188 & \textbf{137 (73\%)} \\ \hline

\multirow{3}{*}{CP}
& AttackLLM & 170 & 12 (7\%) \\ \cline{2-4}
& GENICS    & 175 & 14 (8\%) \\ \cline{2-4}
& \sys      & 171 & \textbf{132 (77\%)} \\ \hline

\multirow{3}{*}{WP}
& AttackLLM & 160 & 11 (7\%) \\ \cline{2-4}
& GENICS    & 165 & 13 (8\%) \\ \cline{2-4}
& \sys      & 158 & \textbf{123 (78\%)} \\ \hline

\multirow{3}{*}{\textbf{Total}}
& AttackLLM & 510 & 35 (7\%) \\ \cline{2-4}
& GENICS    & 525 & 41 (8\%) \\ \cline{2-4}
& \sys      & 517 & \textbf{392 (76\%)} \\ \hline

\end{tabular}
\end{table}

\subsection{\sys vs. Existing Methods (RQ2)}
\label{subsec:rq2}
We compare \sys with existing ICS attack-generation methods to determine whether explicitly reasoning about post-detection recoverability improves the ability to expose TLTR conditions. We consider AttackLLM~\cite{ahmed2025attackllm}, an LLM-based ICS attack-generation approach, and GENICS~\cite{song2024genics}, a knowledge-driven attack-generation framework. These methods provide complementary baselines for evaluating whether TLTR conditions arise systematically without \sys's explicit modeling of recovery constraints and recovery margin.


\shortsectionBf{Methodology.}
For each testbed, we configure AttackLLM, GENICS, and \sys under the same attacker capabilities, manipulable variables, manipulation bounds, and operational constraints. We evaluate the attack scenarios generated by each method in the same vPLC environment used in RQ1. For each execution, attacker access is revoked when detection occurs, after which the same deployed recovery logic is executed. We then apply Algorithm~\ref{alg:recovery} to each execution trace and classify it as \emph{Recoverable}, \emph{Recovery-Infeasible}, or \emph{Recovery-Unsafe}.


\shortsectionBf{Results.}
Table~\ref{tab:rq2_compare} summarizes the comparison. Across the three testbeds, 392 of 517 \sys-generated scenarios (76\%) result in TLTR outcomes. In contrast, only 35 of 510 AttackLLM scenarios (7\%) and 41 of 525 GENICS scenarios (8\%) result in TLTR outcomes. The difference is consistent across all three testbeds: \sys yields TLTR rates of 73--78\%, whereas both baselines remain at 7--8\%. Thus, under the same attacker model and recovery evaluation procedure,
existing attack-generation objectives occasionally produce TLTR
conditions, but they do so far less frequently than \sys's
recovery-aware search.

\shortsectionBf{Analysis.} The difference stems from the objective used to guide scenario generation. AttackLLM and GENICS generate attacks without explicitly targeting the process state from which deployed recovery must operate after detection. Consequently, their generated attacks may achieve process deviations or trigger detection while sufficient recovery margin still remains. In contrast, \sys explicitly incorporates detection conditions, deployed recovery logic, constraint interactions, and constraint-centric process patterns into scenario generation. This directs exploration toward pre-detection states with reduced physical or temporal recovery margin. The results therefore show that TLTR conditions are not merely a byproduct of attack generation: explicitly targeting post-detection recoverability substantially increases their discovery rate.



\begin{tcolorbox}[colback=gray!8,colframe=gray!40,boxrule=0.5pt]
Across the three testbeds, 76\% of \sys-generated scenarios result in
TLTR outcomes, compared with 7\% for AttackLLM and 8\% for GENICS.
Explicitly incorporating post-detection recovery into scenario
generation therefore substantially increases the ability to expose
TLTR conditions.
\end{tcolorbox}

\begin{table}[t!]
\centering
\small
\caption{Effectiveness of ICS defenses against \sys.}
\label{tab:rq3_defense}
\setlength{\tabcolsep}{8pt}
\renewcommand{\arraystretch}{1.08}

\begin{tabular}{|c|c|c|c|}
\hline
\textbf{Testbed} &
\textbf{Defense} &
\textbf{TLTR} &
\textbf{Blocked} \\ \hline\hline

\multirow{2}{*}{MP}
& SAIN  & 137 & 37 (27\%) \\ \cline{2-4}
& Ghani & 137 & 25 (18\%) \\ \hline

\multirow{2}{*}{CP}
& SAIN  & 132 & 41 (31\%) \\ \cline{2-4}
& Ghani & 132 & 29 (22\%) \\ \hline

\multirow{2}{*}{WP}
& SAIN  & 123 & 33 (27\%) \\ \cline{2-4}
& Ghani & 123 & 23 (19\%) \\ \hline

\multirow{2}{*}{\textbf{Total}}
& SAIN  & 392 & 111 (28\%) \\ \cline{2-4}
& Ghani & 392 & 77 (20\%) \\ \hline

\end{tabular}
\end{table}

\subsection{RISK Against ICS Defenses (RQ3)}
\label{subsec:rq3}
We evaluate whether state-of-the-art ICS defenses can detect \sys-generated TLTR attacks early enough to preserve recoverability. We consider the invariant-based detector SAIN~\cite{@SAIN} and the physics-based detector of Ghani et al.~\cite{ghaeini}.

\shortsectionBf{Methodology.}
We evaluate SAIN~\cite{@SAIN} and the physics-based detector of Ghani et al.~\cite{ghaeini} against the \sys-generated TLTR scenarios identified in RQ1. For each scenario, we enable the corresponding defense and repeat the execution under the same testbed configuration. A scenario is considered \emph{blocked} if the defense detects the attack before the process reaches a TLTR state, allowing the deployed recovery logic to remain feasible and safe. If detection occurs only after the process has already entered a TLTR state, the scenario is counted as an unblocked TLTR outcome. The TLTR column in Table~\ref{tab:rq3_defense} reports the \sys-generated TLTR scenarios from RQ1 replayed with each defense enabled.

\shortsectionBf{Results.}
Table~\ref{tab:rq3_defense} summarizes the results. SAIN blocks 111 of 392 TLTR scenarios (28\%), while the physics-based detector blocks 77 of 392 (20\%). The remaining scenarios reach a TLTR state before the defense can preserve recoverability. These results show that detecting malicious behavior is insufficient when detection does not occur before the process crosses the recovery boundary.

\begin{tcolorbox}[colback=gray!8,colframe=gray!40,boxrule=0.5pt]
Overall, state-of-the-art invariant-based and physics-based defenses prevent only a limited fraction of \sys-generated TLTR scenarios. For the remaining scenarios, the defenses do not act early enough to prevent the system from reaching a state where recovery is infeasible or unsafe. These results highlight a limitation of existing ICS defenses: detecting anomalous behavior does not ensure that detection occurs early enough to preserve recovery feasibility.
\end{tcolorbox}

\subsection{RISK Sensitivity (RQ4)}
\label{subsec:rq4}

We evaluate how sensitive \sys is to variations in operational traces and automatically generated candidate attack scripts. We conduct two complementary sensitivity tests. 

\shortsectionBf{(1) Sensitivity to trace availability.} To evaluate robustness to incomplete operational data, we construct multiple trace subsets for each testbed by randomly sampling 60\%, 70\%, 80\%, and 90\% of the original operational logs. For each subset, we re-run constraint-centric pattern extraction and regenerate candidate attack scripts using the resulting empirical bounds. For each configuration, we generate and evaluate 100 attack trajectories under identical attacker capabilities, validation procedures, and recovery analysis.

Across all testbeds, reducing trace coverage has only a limited impact on \sys's ability to identify TLTR scenarios. When using 60\% of the available traces, \sys achieves TLTR rates in the range of 72\% to 75\%. With 80\% or more traces, performance remains within 3--5\% of the full-trace baseline. This indicates that \sys does not rely on exhaustive operational logs; a moderate amount of trace data is sufficient to maintain stable TLTR identification.

\shortsectionBf{(2) Sensitivity to attack script generation.}
\sys uses an off-the-shelf LLM as a structured proposal mechanism for proposing candidate attack scripts under a fixed schema. Since the LLM is not used for feasibility reasoning, we evaluate robustness by varying prompt formulations and independently regenerating candidate scripts.

Specifically, for each testbed, we construct four semantically equivalent prompt variants that describe the same attack objective, target constraints, manipulable variables, and admissible ranges. The variants differ only in how this information is presented (e.g., reordered fields, emphasis on timing versus magnitude, inclusion of short examples, or alternative formatting). For each variant, we independently generate multiple sets of candidate attack scripts and validate them under the same constraint-derived bounds. All generated scripts are subjected to identical constraint-based validation and vPLC execution to assess consistency in TLTR identification.


Different prompt formulations produce diverse script structures, including variations in trigger thresholds, manipulation timing, and variable combinations. However, after constraint-based validation, most scripts converge to similar admissible manipulation patterns. Across all prompt variants, the TLTR rate varies by at most 6\%, and in all configurations the rate remains around 75\% to 80\%. These results show that surface-level variability in attack script generation does not significantly affect \sys's ability to identify TLTR scenarios.


\begin{tcolorbox}[colback=gray!8,colframe=gray!40,boxrule=0.5pt]
Overall, \sys is robust to incomplete trace data and variations in automatic attack script generation. Its effectiveness in identifying TLTR conditions is primarily determined by the extracted constraint model and empirical recovery bounds, rather than by specific trace realizations or attack script instances. This robustness supports the reliability and reproducibility of \sys in realistic deployment settings.
\end{tcolorbox}

\begin{table}[t!]
\centering
\small
\caption{\sys applicability across ICS platforms.}
\label{tab:rq5_generality}
\setlength{\tabcolsep}{8pt}
\begin{threeparttable}
\begin{tabular}{|p{4.5cm}|c|c|c|}
\hline
\textbf{Pipeline Component} &
\textbf{MP} &
\textbf{CP} &
\textbf{WP} \\ \hline\hline

PLC constraint extraction & \checkmark & \checkmark & \checkmark \\ \hline
Operator guideline parsing & \checkmark & \checkmark & \checkmark \\ \hline
Multi-layer model construction & \checkmark & \checkmark & \checkmark \\ \hline
Constraint-centric pattern extraction & \checkmark & \checkmark & \checkmark \\ \hline
Attack script generation & \checkmark & \checkmark & \checkmark \\ \hline
vPLC-based execution & \checkmark & \checkmark & \checkmark \\ \hline
Recovery feasibility analysis & \checkmark & \checkmark & \checkmark \\ \hline
Platform-specific tuning required & No & No & No \\ \hline

\end{tabular}
\begin{tablenotes}
\footnotesize
\item MP: Manufacturing plant; CP: Chemical plant; WP: Water plant.
\end{tablenotes}
\end{threeparttable}
\end{table}

\subsection{\sys Generality (RQ5)}
\label{subsec:rq5}
We evaluate the generality of \sys across three ICS platforms with different PLC vendors, programming environments, and industrial processes: a manufacturing system using a Siemens S7-1500, a chemical processing system using a PLCNext controller, and a water treatment system using a CODESYS PLC. For each testbed, we apply the same analysis workflow, including constraint extraction, pattern analysis, attack script generation, vPLC-based execution, and recovery feasibility analysis.

Table~\ref{tab:rq5_generality} shows that all components of the \sys pipeline operate across the three platforms without platform-specific retuning. Moreover, \sys identifies TLTR scenarios at comparable rates across the evaluated systems: 73\% in MP, 77\% in CP, and 78\% in WP (Table~\ref{tab:rq1_original}). These results show that the workflow can be applied across different PLC vendors, programming environments, and process domains.

\sys is applicable to ICS deployments in which recovery-relevant control logic, detection policies, recovery procedures, and operational behavior are available for analysis. It is not directly applicable when post-detection behavior depends solely on mechanisms outside this model, such as purely mechanical protection devices, or when proprietary controllers do not expose the information required to reconstruct recovery-relevant semantics.

\section{Industry Experience: Fertilizer Plant Case}
\label{sec:industry}
To assess \sys real-world applicability, we applied \sys to a commercial fertilizer plant, focusing on the steam boiler subsystem. For this subsystem, we obtained the corresponding PLC program, (machine-parsable) operator recovery guidelines, and relevant safety and detection logic, together with 12 days of operational traces for subsystem-specific variables.

\shortsectionBf{Offline analysis and virtual execution.}
All analysis in this case study was conducted offline. \sys did not interact with or issue commands to the live production system. Instead, the provided PLC programs, safety logic, and operational traces were integrated into a virtual PLC environment and a subsystem-level simulator capable of replaying real plant behaviors. This setup enabled realistic recovery analysis without affecting the plant's production operations.

\shortsectionBf{Industrial process.}
The steam boiler maintains the steam drum level ($L_1$) by regulating the burner firing rate ($F_{fuel}$) and the feedwater control valve ($FCV$). When the deployed detection system determines that the drum level has crossed the alarm threshold, the recovery mechanism reduces boiler firing and increases feedwater flow to stabilize the process. If the drum level cannot be stabilized and reaches the emergency-trip threshold, the boiler protection system automatically shuts down the unit.

\shortsectionBf{Identified TLTR scenario.}
Using the extracted multi-layer constraint model and operational traces, \sys identified a TLTR scenario in which adversarial manipulations progressively reduced the process's recoverability while remaining within all configured operating limits. The attacker gradually increased the burner firing rate from 60\% to 80\%, causing the steam drum water level to gradually decline while remaining within the acceptable operating range. As the drum level approached the  alarm threshold, the attacker initiated the closing of motor-operated feed water valve, sharply reducing the feed water inflow and accelerating the decline of the steam drum water level.


Once the drum level crossed the alarm threshold, the detection system correctly detected the attack, immediately revoked the attacker's remote access, and activated the recovery mechanism. The recovery mechanism responded exactly as specified by reducing boiler firing and increasing feedwater flow according to the deployed recovery procedure.

However, the steam drum level continued to decrease. Although recovery was initiated after detection, the remaining recovery margin was insufficient for the available recovery actions to prevent the steam drum level from reaching the emergency-trip threshold. Consequently, the boiler protection system initiated an emergency shutdown that prevented further process degradation but forced an unplanned interruption of plant operations, representing escalation to the protection layer rather than successful completion of the deployed recovery procedure.

\sys classified the scenario as \emph{recovery-infeasible} because the prescribed recovery actions, although initiated immediately after detection, were unable to reach their intended recovery condition before the emergency-trip threshold was reached. The attack neither bypassed the detection system nor interfered with the recovery mechanism. Instead, the adversarial manipulations progressively drove the process beyond the point at which the deployed recovery procedure could prevent the emergency shutdown.



\shortsectionBf{Operational feedback and impact.}
We presented this margin-exhaustion behavior and its quantitative implications to plant operators and engineers, who confirmed its operational relevance. Based on this feedback, the plant updated its recovery guidance to incorporate earlier intervention triggers, explicit verification of valve responsiveness, revised controller tuning procedures, and periodic inspection and calibration of recovery-critical actuators to reduce the likelihood of TLTR conditions.

This case study was conducted with operator consent and cooperation, and sensitive operational details were anonymized to preserve confidentiality.

\section{Limitations And Discussions}


\shortsectionBf{Human-in-the-loop recovery adaptation.}
\sys executes deployed recovery procedures to ensure reproducible and conservative evaluation, but does not explicitly model human improvisation, experience-based adaptations, or collaborative decision-making during real incidents. In practice, although operators may deviate from prescribed guidelines, such adaptations remain fundamentally constrained by bounded actuation rates, safety interlocks, and time-critical dynamics once recovery margins are severely reduced. Incorporating controlled human-in-the-loop studies and field observations is an important direction for future work.

\shortsectionBf{Attacker knowledge and exploitability.}
\sys audits TLTR vulnerabilities from the operator's perspective using deployment-specific artifacts available to the asset owner. An attacker, however, may have only partial or approximate knowledge of the target process, detection conditions, and recovery behavior. Such uncertainty may affect the attacker's ability to drive the process into a low-recovery-margin state, but does not change whether that state is recoverable once reached. \sys therefore evaluates the recoverability of the deployed detection--recovery configuration under admissible attacker manipulations, rather than assuming that the attacker possesses the same system knowledge available to the operator.

\shortsectionBf{Role and limitations of LLM-guided generation.}
\sys employs large language models only as structured template proposal mechanisms, while all physical and logical feasibility guarantees are enforced through constraint validation. The diversity of generated candidate attack scripts depends on prompt design and model behavior. Our sensitivity analysis indicates limited impact on the final TLTR outcomes. While alternative approaches such as optimization or reachability-based search could be used, they require precise system models and explicit objective formulations over recovery failure conditions, which are not available in our setting due to partial observability and cross-layer interactions. \sys therefore adopts LLM-guided generation as a practical mechanism for structured exploration under such constraints.

\shortsectionBf{Coverage and completeness.}
\sys does not guarantee exhaustive discovery of all recovery failure scenarios. It should be viewed as a systematic stress-testing approach that explores constraint-admissible trajectories to uncover diverse TLTR attack scenarios within practical limits, rather than a complete verification method. This analysis relies on recovery-relevant behaviors derived from control logic, system functionality, and observed operational patterns. While this captures a broad range of admissible recovery actions in practice, settings with informal or undocumented recovery procedures may introduce additional variability not fully reflected in the extracted model.

\shortsectionBf{Mitigation Strategies.} \label{sec:defense} Defending against TLTR vulnerabilities requires jointly designing the detection and recovery mechanisms rather than considering them independently. Advancing alarm generation may preserve additional recovery time, but overly aggressive detection policies can increase false positives. Conversely, improving detection alone may not be sufficient if the remaining time and control authority available after detection cannot support the documented recovery procedures. Defenses should therefore configure alarm thresholds using both detection objectives and recoverability constraints. Offline analysis using \sys\ can identify TLTR conditions prior to deployment, enabling refinement of alarm thresholds, control logic, and recovery policies.


\section{Related Work} Prior ICS security research has studied attack detection and prevention~\cite{ghaeini,vetplc,scaphy,abdelaty2021daics,@SAIN,yilmaz2018attack, wolsing2025gecos}, recovery planning~\cite{staves2020framework,he2015industrial,alves2020secure, powell2026responding}, and adversarial testing and attack generation~\cite{song2024genics,ahmed2025attackllm,das2020anomaly,umer2021attack, villa2025icsquartz, cheng2026llm}. Detection and prevention approaches aim to identify or restrict adversarial behavior, but do not explicitly evaluate whether sufficient recovery margin remains when detection occurs. Recovery-planning approaches design procedures to restore safe operation after disruptions, generally considering recovery under modeled operating conditions rather than adversarially reduced recovery margins. Attack-generation approaches primarily search for manipulations that cause physical or operational impact, while stealthy attacks pursue such impact without being detected~\cite{urbina2016limiting}; although these attacks may incidentally reduce recovery margin, they do not explicitly target states that are no longer recoverable at detection. In contrast, \sys  \emph{jointly audits a subject ICS' detection and recovery} by identifying TLTR scenarios in which adversarial manipulations drain the available recovery margin before detection, such that the deployed recovery procedure no longer has sufficient margin to succeed. Appendix~\ref{lab:relatedwork} provides a detailed comparison with the related work.

\section{Conclusion}
We study post-detection recoverability in industrial control systems and formulate too-late-to-recover (TLTR) vulnerabilities, where adversarial manipulations drain recovery margin before detection. We present \sys, an automated framework for discovering and validating TLTR scenarios by jointly auditing detection and recovery. Across three industrial testbeds and a real-world fertilizer plant, our results show that successful detection does not necessarily imply successful recovery, motivating recoverability at detection as an explicit ICS security objective.


\newpage
\section*{Ethics Considerations}
\sys is intended for defensive auditing of post-detection recoverability. It explores admissible pre-detection manipulations but does not introduce new mechanisms for unauthorized access, bypassing detection, or disabling recovery. All adversarial actions terminate at detection.

Experiments were conducted in controlled evaluation environments, including vPLCs and dedicated physical/high-fidelity testbeds. The fertilizer-plant case study was performed offline using plant-provided artifacts and operational traces; no attacks were executed against the live production system. Plant operators and engineers were involved in the study, sensitive operational details were anonymized, and the identified recovery limitation informed updates to recovery guidance.


\bibliographystyle{IEEEtran}
\bibliography{reference}

\appendices

\begin{table*}[t]
\centering
\small
\caption{Comparison of \sys with relevant systems.
}
\label{tab:comparison}

\begin{tabular}{|l|c|c|c|c|c|}
\hline
\textbf{System}
& \textbf{Attack Detection}
& \textbf{Prevention}
& \textbf{Recovery Planning}
& \textbf{Attack Generation}
& \textbf{TLTR Recoverability Analysis} \\
\hline

Ghani et al.~\cite{ghaeini}
& \checkmark & \xmark & \xmark & \xmark & \xmark \\
\hline

VETPLC~\cite{vetplc}
& \checkmark & \xmark & \xmark & \xmark & \xmark \\
\hline

SCAPHY~\cite{scaphy}
& \checkmark & \xmark & \xmark & \xmark & \xmark \\
\hline

DAICS~\cite{abdelaty2021daics}
& \checkmark & \xmark & \xmark & \xmark & \xmark \\
\hline

SAIN~\cite{@SAIN}
& \checkmark & \checkmark & \xmark & \xmark & \xmark \\
\hline

Yılmaz et al.~\cite{yilmaz2018attack}
& \xmark & \checkmark & \xmark & \xmark & \xmark \\
\hline

GENICS~\cite{song2024genics}
& \xmark & \xmark & \xmark & \checkmark & \xmark \\
\hline

AttackLLM~\cite{ahmed2025attackllm}
& \xmark & \xmark & \xmark & \checkmark & \xmark \\
\hline

AttackRules~\cite{umer2021attack}
& \xmark & \xmark & \xmark & \checkmark & \xmark \\
\hline

Staves et al.~\cite{staves2020framework}
& \xmark & \xmark & \checkmark & \xmark & \xmark \\
\hline

He et al.~\cite{he2015industrial}
& \xmark & \xmark & \checkmark & \xmark & \xmark \\
\hline

Alves et al.~\cite{alves2020secure}
& \xmark & \xmark & \checkmark & \xmark & \xmark \\
\hline

\sys
& \xmark & \xmark & \xmark & \checkmark & \checkmark \\
\hline

\end{tabular}
\end{table*}

\section{Detailed Related-Work Comparison}
\label{lab:relatedwork}
We compare \sys with prior work in Table~\ref{tab:comparison}.

\shortsectionBf{Attack detection.}
Prior work on anomaly and attack detection in ICS focuses on identifying deviations from expected process behavior using physical models, control invariants, and data-driven techniques~\cite{ghaeini,vetplc,scaphy,abdelaty2021daics}. These systems aim to raise timely alarms and trigger recovery, implicitly assuming that detection occurs early enough for safe recovery. In contrast, \sys audits whether the process remains recoverable at detection and identifies TLTR scenarios in which recovery becomes infeasible or unsafe despite correct detection.

\shortsectionBf{Prevention.} Prevention-oriented approaches seek to block or constrain malicious actions through access control, network segmentation, authentication, and safety interlocks~\cite{@SAIN,yilmaz2018attack, rasapour2019framework}. These mechanisms reduce the attack surface and restrict adversarial manipulations, but do not assess whether the process remains recoverable once an attack has occurred and been detected. In contrast, \sys audits post-detection recoverability by identifying TLTR scenarios in which the deployed recovery actions have become infeasible or unsafe.


\shortsectionBf{Recovery planning.}
Recovery planning frameworks design procedures and control strategies to restore the process to a safe operating state after disruptions~\cite{staves2020framework,he2015industrial}. These approaches assume that recovery actions remain executable and safe under modeled conditions. \sys instead challenges this assumption by identifying TLTR scenarios in which deployed recovery actions become infeasible or unsafe under system constraints.

\shortsectionBf{Attack generation.} Prior work on adversarial testing and attack generation aims to discover attacks that achieve physical disruptions or safety violations~\cite{song2024genics,ahmed2025attackllm,das2020anomaly,umer2021attack}. These approaches therefore search for manipulations that drive the process toward unsafe states or physical impact. Stealthy ICS attacks similarly seek to achieve physical impact while evading detection~\cite{urbina2016limiting}. Although such attacks may incidentally leave insufficient recovery margin after detection, prior work does not explicitly model post-detection recoverability or systematically search for states from which the deployed recovery actions can no longer restore the process to a safe operating state. In contrast, \sys explicitly searches for adversarial scenarios that drive the process into states that are no longer recoverable at detection, while leaving the deployed detection and recovery mechanisms operational. Thus, an attack may be correctly detected and recovery may proceed as intended, yet the process may already be too late to recover. 

\begin{tcolorbox}[colback=gray!10,colframe=gray!40,boxrule=0.3pt]
\noindent
\sys complements existing ICS security approaches by enabling operators and engineers to audit post-detection recoverability and proactively identify TLTR scenarios in which deployed recovery actions become infeasible or unsafe.
\end{tcolorbox}

\begin{table*}[t]
\centering
\scriptsize
\caption{TLTR scenarios identified by \sys under the original ICS configurations.}
\label{tab:tltr_examples}
\renewcommand{\arraystretch}{1.15}
\setlength{\tabcolsep}{2.8pt}

\begin{threeparttable}
\begin{tabular}{
|p{0.62cm}
|p{0.78cm}
|p{3.05cm}
|p{2.75cm}
|p{3.05cm}
|p{5.20cm}|}
\hline
\textbf{ID} &
\textbf{Type} &
\textbf{Pre-detection Manipulation} &
\textbf{State at Detection} &
\textbf{Deployed Recovery $\rho$} &
\textbf{TLTR Evidence and Outcome} \\
\hline\hline

\multicolumn{6}{|l|}{\textbf{Manufacturing Plant (MP)}} \\
\hline

MP-4 &
Inf. &
Advance the VGR gripper-release timing by approximately $0.4$\,s
during payload transfer. &
At $t_d=63.2$\,s, release occurs approximately $0.4$\,s early and
the workpiece remains about $12$\,mm above the hand-off position. &
Restore the nominal release timing and pause the transfer axis using
the deployed bounded-deceleration procedure. &
The transfer axis requires approximately $0.75$\,s to arrest, whereas
the workpiece leaves the recoverable gripper/hand-off envelope in
approximately $0.4$\,s. The required recovery time therefore exceeds
the remaining physical margin by approximately $0.35$\,s, and the
workpiece leaves the envelope at $t_v\approx63.6$\,s before recovery
completes.
\\
\hline

MP-6 &
Inf. &
Increase the loaded high-bay crane's vertical travel speed from
approximately $0.30$ to $0.42$\,m/s while approaching the storage
rack. &
At $t_d=141.0$\,s, the crane travels at $0.42$\,m/s with
approximately $35$\,mm remaining before the shelf. &
Apply the deployed bounded-deceleration stop with maximum
deceleration of approximately $4$\,m/s$^2$. &
Reaction-delay travel is approximately $21$\,mm and subsequent
stopping distance approximately $22$\,mm, requiring about $43$\,mm
in total. Only $35$\,mm remains at detection. Thus, the deployed stop
cannot prevent contact, and the crane reaches the shelf before coming
to rest.
\\
\hline

MP-7 &
Uns. &
Reduce the pneumatic gripper supply pressure from approximately
$6.0$ to $4.2$\,bar. &
At $t_d=55.0$\,s, the cylinder is mid-grip and only a small
gripping-force margin remains. &
Restore pneumatic pressure to the nominal $6$\,bar setting and
re-establish the grip. &
Pressure restoration produces a transient peak of approximately
$7.1$\,bar due to pneumatic-line dynamics, exceeding the
approximately $6.8$\,bar-equivalent workpiece crush boundary.
The secondary constraint is satisfied at detection but is violated
after recovery begins, causing the workpiece to crack before pressure
settles.
\\
\hline

MP-8 &
Inf. &
Bias the MPO turntable index position to approximately
$2.3^\circ$ of misalignment. &
At $t_d=97.0$\,s, the turntable is misaligned by approximately
$2.3^\circ$ while the VGR pick motion is approaching the interference
region. &
Execute the deployed full-rotation turntable homing procedure. &
Homing requires approximately $1.8$\,s, whereas the already-committed
VGR pick motion reaches the interference region in approximately
$0.4$\,s. The VGR therefore reaches the turntable before homing can
restore alignment.
\\
\hline

\multicolumn{6}{|l|}{\textbf{Chemical Processing Plant (CP)}} \\
\hline

CP-7 &
Uns. &
Increase catalyst feed, raising the reactor exotherm while the
downstream neutralization system remains within its admissible
operating region. &
At $t_d\approx360$\,s, the reactor exotherm is elevated while the
downstream treatment load remains within its capacity. &
Reduce catalyst feed and increase the deployed quench flow to recover
the reactor temperature. &
The required quench flow increases downstream effluent loading.
The neutralization subsystem reaches its throughput/overflow boundary
approximately $105$\,s after recovery starts, whereas reactor recovery
requires approximately $150$\,s. Thus, $t_d<t_v<t_c$: the secondary
violation occurs during correct recovery and before primary recovery
completes.
\\
\hline

CP-8 &
Inf. &
Bias the heat-exchanger coolant control to reduce cooling
effectiveness. &
At $t_d=400$\,s, reactor temperature is approximately
$148^{\circ}$C, the trip boundary is $150^{\circ}$C, and temperature
is rising at approximately $0.5^{\circ}$C/s. &
Command the coolant valve to maximum aperture under the deployed
water-hammer ramp limiter. &
The valve requires approximately $2.67$\,s to reach full aperture,
followed by additional thermal transport delay. The approximately
$2^{\circ}$C margin at detection is exhausted before effective cooling
is established, causing temperature to cross the $150^{\circ}$C
boundary before recovery can take effect.
\\
\hline

CP-9 &
Uns. &
Increase the pressure-relief back-pressure setting within its
configured range. &
At $t_d=480$\,s, reactor pressure is approximately $11.2$\,bar
against a $12.0$\,bar trip boundary, while the downstream flare-header
constraint remains satisfied. &
Open the deployed relief vent fully to reduce reactor pressure. &
Full recovery venting produces approximately $18$\,kg/s discharge
into a flare header designed for approximately $15$\,kg/s. The
downstream over-pressure condition occurs about $12$\,s after recovery
begins, whereas reactor pressure requires about $40$\,s to reach its
recovery target. The recovery action therefore causes a secondary
violation before recovery completes.
\\
\hline

\multicolumn{6}{|l|}{\textbf{Water Treatment Plant (WP)}} \\
\hline

WP-6 &
Inf. &
Reduce the chemical-dosing rate while raw-water flow remains nominal,
allowing under-treated water to pass the dosing point. &
At $t_d=340$\,s, approximately $45$\,L of under-treated water has
already entered the downstream UF-stage piping. &
Increase chemical dosing to the deployed recovery setting while
maintaining downstream flow. &
The affected water is already downstream of the actuation point.
At approximately $2.5$\,L/s, it reaches the downstream stage after
about $18$\,s, whereas the corrected dose requires approximately
$25$\,s to reach the UF inlet. The deployed recovery therefore cannot
affect the already-transited volume before it reaches the downstream
boundary.
\\
\hline

WP-7 &
Uns. &
Increase the RO feed-pressure setpoint within its admissible range. &
At $t_d=210$\,s, membrane differential pressure is approximately
$1.85$\,bar, below the $2.3$\,bar mechanical boundary. &
Reduce RO feed pressure and initiate the deployed membrane flush
cycle. &
The flush produces a temporary flow transient of approximately
$1.4\times$ nominal flow while the elevated differential pressure has
not yet decayed. Differential pressure reaches approximately
$2.35$\,bar at $t_v\approx t_d+4$\,s, exceeding the $2.3$\,bar
secondary mechanical boundary before the flush completes at
$t_c\approx t_d+8$\,s.
\\
\hline

WP-9 &
Uns. &
Reduce sodium-bisulfite dosing while the downstream RO influent
remains within its admissible quality constraint. &
At $t_d=150$\,s, residual chlorine is near its detection boundary
while the RO-influent secondary constraint remains satisfied. &
Increase bisulfite dosing to the deployed recovery setting to
neutralize the residual chlorine. &
The recovery command produces a short dosing transient above the
downstream oxidant-scavenger limit. The secondary RO-influent
constraint is violated at approximately $t_d+4$\,s, whereas
restoration of the primary chlorine condition requires approximately
$20$\,s. Thus, the secondary violation occurs after recovery begins
and before primary recovery completes.
\\
\hline

\end{tabular}

\begin{tablenotes}[flushleft]
\scriptsize
\item Inf.: recovery-infeasible; Uns.: recovery-unsafe.
For all scenarios, adversarial manipulation terminates at detection;
the reported violation occurs during execution of the deployed
recovery logic and before its specified recovery condition is reached.
\end{tablenotes}

\end{threeparttable}
\end{table*}

\section{TLTR Scenarios Identified by \sys}
\label{app:tltr_examples}

Table~\ref{tab:tltr_examples} presents ten TLTR scenarios identified by
\sys under the original ICS configurations. The scenarios span the
manufacturing, chemical processing, and water treatment testbeds and
include both recovery-infeasible and recovery-unsafe outcomes. For each
scenario, we report the pre-detection manipulation, the process state
at detection, the deployed recovery procedure, and the condition that
prevents recovery from completing safely. In every scenario, attacker
manipulation terminates at detection.

\section{LLM Prompt Templates}
\label{app:llm_prompts}

This appendix describes the prompt structure used by \sys for
LLM-assisted proposal of candidate TLTR attack scripts. The LLM is used
only as a structured proposal mechanism. It receives the multi-layer
constraint model (MCM) and constraint-centric patterns extracted from
operational traces and proposes candidate manipulations, trigger
conditions, and approximate parameters. \sys subsequently validates all
proposed candidates against the attacker model and constraint-derived
bounds, executes them in the vPLC environment, and determines their
recoverability outcome.

\subsection{Prompt Structure}

Each query consists of a system prompt that constrains the role of the
LLM and a user prompt containing system-specific inputs derived from the
MCM and trace analysis.

\paragraph{System Prompt.}
All queries use the following system prompt:

\begin{quote}
You are a structured proposal mechanism for an offline industrial
control system recovery stress-testing framework.

Your task is to propose candidate attack scripts that reduce the
physical or temporal recovery margin before detection.

Follow these rules:

1) Output valid JSON only.

2) Follow the required output schema exactly.

3) Use only the provided attacker-accessible variables and their
admissible manipulation bounds.

4) Respect all provided control, detection, recovery, protection, and
actuation constraints.

5) Do not introduce new variables, sensors, actuators, alarms,
constraints, or recovery actions.

6) Do not modify PLC control logic, detection mechanisms, recovery
procedures, or protection mechanisms.

7) Specify manipulations as condition-triggered stages, including the
variable, approximate magnitude, and duration where applicable.

8) All adversarial manipulation must terminate when the specified
detection condition occurs. After detection, propose no additional
attacker action.

9) Do not claim that a proposed script is feasible, successful, or
too-late-to-recover. Feasibility and recoverability are determined by
the external validation and execution system.

10) Keep all proposed manipulations within the supplied admissible
bounds.
\end{quote}

\paragraph{User Prompt.}
For each target ICS, \sys supplies the following system-specific
information:

\begin{itemize}
    \item \textbf{Target system:} Process description and relevant
    process variables.

    \item \textbf{Attacker-accessible variables:} Variables that the
    attacker may manipulate, together with their types and admissible
    manipulation ranges.

    \item \textbf{Control constraints:} PLC conditions, actions,
    actuator limits, interlocks, and rate constraints relevant to the
    target variables.

    \item \textbf{Detection conditions:} Alarm predicates and the
    process conditions that trigger detection.

    \item \textbf{Recovery logic:} Deployed recovery actions,
    triggering conditions, actuation limits, and recovery completion
    conditions.

    \item \textbf{Protection constraints:} Available protection
    mechanisms and triggering conditions, when present.

    \item \textbf{Constraint-centric patterns:} Trace-derived
    low-margin process regions characterized by the remaining slack
    $D_i$, estimated time to the relevant boundary $T_i$, recovery
    rate $R_i$, and associated process states.

    \item \textbf{Attack objective:} Propose a candidate script that
    drives the process toward a low-recovery-margin state before
    detection while satisfying the supplied attacker and system
    constraints.
\end{itemize}

To explore both TLTR failure modes, \sys instantiates the attack
objective in two forms:

\begin{quote}
\textbf{Recovery-infeasible objective:}
Propose a pre-detection manipulation sequence that reduces the
remaining recovery margin such that, after detection and termination of
the attack, the deployed recovery logic may be unable to reach its
specified recovery condition before the relevant target constraint is
violated or the recovery interval expires.

\textbf{Recovery-unsafe objective:}
Propose a pre-detection manipulation sequence that places the process
in a state where, after detection and termination of the attack,
execution of the deployed recovery logic may cause a
recovery-associated secondary constraint to be violated before the
primary recovery condition is reached.
\end{quote}

The LLM is not instructed to determine whether either condition
actually occurs. These objectives only guide candidate proposal; \sys
determines the outcome from vPLC execution and
Algorithm~\ref{alg:recovery}.

\subsection{Output Schema}

The LLM output follows the JSON structure below:

\begin{quote}
\{
  "attack\_script": [
    \{
      "stage": "{stage name}",
      "condition": "{trigger condition}",
      "action": "{manipulation within admissible bounds}",
      "duration": "{time- or condition-based duration}"
    \}
  ],
  "observation\_horizon": "{post-detection observation window}"
\}
\end{quote}

The stages specify only pre-detection adversarial manipulations.
Execution of the attack script terminates when detection occurs, after
which \sys observes the process for the specified horizon $H$ while the
deployed control and recovery logic executes without further adversarial
manipulation.


\end{document}